\documentclass[%
 aip,
 jap,
 amsmath,amssymb,
 reprint,%
floatfix,
]{revtex4-1}

\usepackage{float}
\usepackage{graphicx}
\usepackage{dcolumn}
\usepackage{bm}

\usepackage[utf8]{inputenc}
\usepackage[T1]{fontenc}
\usepackage{mathptmx}
\usepackage{etoolbox}
\usepackage{cancel}
\usepackage{xcolor}
\usepackage{soul}
\usepackage{booktabs}
\usepackage{orcidlink}

\makeatletter
\def\@email#1#2{%
 \endgroup
 \patchcmd{\titleblock@produce}
  {\frontmatter@RRAPformat}
  {\frontmatter@RRAPformat{\produce@RRAP{*#1\href{mailto:#2}{#2}}}\frontmatter@RRAPformat}
  {}{}
}%
\makeatother

\newcommand{\bl}[1]{{\textbf{#1}}}
\newcommand{\del}[2]{\frac{\partial{#1}}{\partial{#2}}}
\newcommand{\bra}{\ensuremath{\langle}}
\newcommand{\ket}{\ensuremath{\rangle}}

\usepackage[normalem]{ulem}

\begin{document}

\preprint{AIP/123-QED}

\title{Tutorial on Superconducting Quantum Circuits: From Basics to Applications\\}
\author{Denys Derlian Carvalho Brito (corresponding author)~\orcidlink{0009-0007-1669-0340}}
\email{derlian@ita.com}
\affiliation{ 
Departamento de Fisica, Instituto Tecnológico de Aeronáutica, Praça Marechal Eduardo Gomes, São José dos Campos, 12228-900, São Paulo, Brasil.
}%

 \author{Fernando Valadares~\orcidlink{0000-0002-7961-2215}}
\email{fernando.valadares@u.nus.edu}
\affiliation{ 
Centre for Quantum Technologies, National University of Singapore, Lower Kent Ridge Rd, Queenstown, 119077, Singapore
}%

\author{André Jorge Carvalho Chaves (corresponding author)~\orcidlink{0000-0003-1381-8568}}
 \email{andrejck@ita.br}
\affiliation{ 
Departamento de Fisica, Instituto Tecnológico de Aeronáutica, Praça Marechal Eduardo Gomes, São José dos Campos, 12228-900, São Paulo, Brasil.
}%

\date{\today}

\begin{abstract}
As superconducting circuits emerge as a leading platform for scalable quantum information processing, building comprehensive bridges from the foundational principles of macroscopic quantum phenomena to the architecture of modern quantum devices is increasingly essential for introducing new researchers to the field. This tutorial provides a self-contained, pedagogical introduction to superconducting quantum circuits at the undergraduate level. Beginning with an overview of superconductivity and the Josephson effect, the tutorial systematically develops the quantization of microwave circuits into the framework of circuit quantum electrodynamics (cQED). The transmon qubit is then introduced as a state-of-the-art application, with a detailed derivation of its Hamiltonian and its interaction with control and readout circuitry. The theoretical formalism is consolidated through a numerical simulation of vacuum Rabi oscillations in a driven transmon-resonator system, a canonical experiment that demonstrates the coherent energy exchange characteristic of the strong coupling regime. This work serves as a foundational guide and first point of contact, equipping students and researchers with the conceptual and mathematical tools necessary to understand and engineer superconducting quantum hardware. 
\end{abstract}

\maketitle

\tableofcontents

\section{Introduction}\label{sec1}

Quantum technologies, including quantum computing, quantum sensing, and quantum communication, are at the forefront of what is commonly referred to as the “Second Quantum Revolution''~\cite{Pasquazi2024}. This emerging paradigm arises from the intersection of quantum mechanics and information science, which forms the foundation of Quantum Information Science (QIS)~\cite{Paul_2018}. Quantum computing, in particular, has garnered significant attention due to its potential to solve certain computational problems exponentially faster than classical approaches~\cite{dalzell2023quantumalgorithmssurveyapplications, oliveira2024}. Applications of this technology span numerous domains, including scientific research, industrial innovation, national security, and real-time data processing~\cite{dalzell2023quantumalgorithmssurveyapplications, Kumar_2024, Thomas_2024}.

Among the leading physical implementations of quantum processors, superconducting qubits have emerged as a highly promising platform~\cite{Plourde_2022, kjaergaard_superconducting_2020}. These systems offer key advantages such as compatibility with microwave control technologies, integration with existing lithographic and electronic techniques, and scalability to multi-qubit architectures~\cite{Chen_2023, Chohan_2024}. Their operation at cryogenic temperatures ensures low energy dissipation and extended coherence times, both essential for the reliable execution of quantum algorithms and the realization of fault-tolerant quantum computing~\cite{https://doi.org/10.48550/arxiv.2409.02063}.

Given their unique properties, superconducting qubits are positioned as a cornerstone of emerging quantum computing architectures. Their utility extends beyond computational applications to encompass secure communication systems and advanced sensing platforms~\cite{Thomas_2022, Koch_2022}. A comprehensive understanding of their physical principles, design considerations, and control mechanisms is thus essential for researchers and engineers engaged in the development of practical quantum technologies.

However, for students and researchers entering the field, the path from the underlying physics to a functional quantum device can be daunting, as it requires synthesizing concepts from condensed matter physics, quantum optics, and microwave engineering. A clear, pedagogical bridge from first principles to modern hardware is therefore essential for training the next generation of quantum engineers and scientists. This tutorial provides this bridge, starting from the basics of superconductivity to the introductory design and operation of a state-of-the-art transmon qubit. This tutorial is organized as follows: In Sec. \ref{sec:superconductivity}, we discuss the basics of superconductivity, including the Meissner effect and the London equation. In Sec. \ref{sec:josephson}, the Josephson effect is introduced; in Sec. \ref{sec:CQE}, we present some aspects of circuit quantum electrodynamics, which allow the definition of the transmon in Sec. \ref{sec:transmon}. Finally, in Sec. \ref{sec:app} some applications are presented.

\section{Superconductivity} \label{sec:superconductivity}
The liquefaction of Helium, performed by H. Kamerlingh Onnes in 1908~\cite{Onnes1991}, set the grounds for the investigation of physics at few-Kelvin temperatures, leading to the development of many cryogenic technologies in the subsequent decades~\cite{cern2023cryogenics}. In 1911, the phenomenon of superconductivity was observed for the first time with the abrupt fall of mercury's resistivity below a \textit{critical temperature} ($T_c$) of approximately 4.1\,K~\cite{timm2023theory}. Although mercury below a specific critical temperature $T_c$ was initially assumed to be a perfect conductor, it was later demonstrated that this is not the case for a superconductor~\cite{henyey1982}.
\begin{figure}[!ht]
    \includegraphics[width = 76mm]{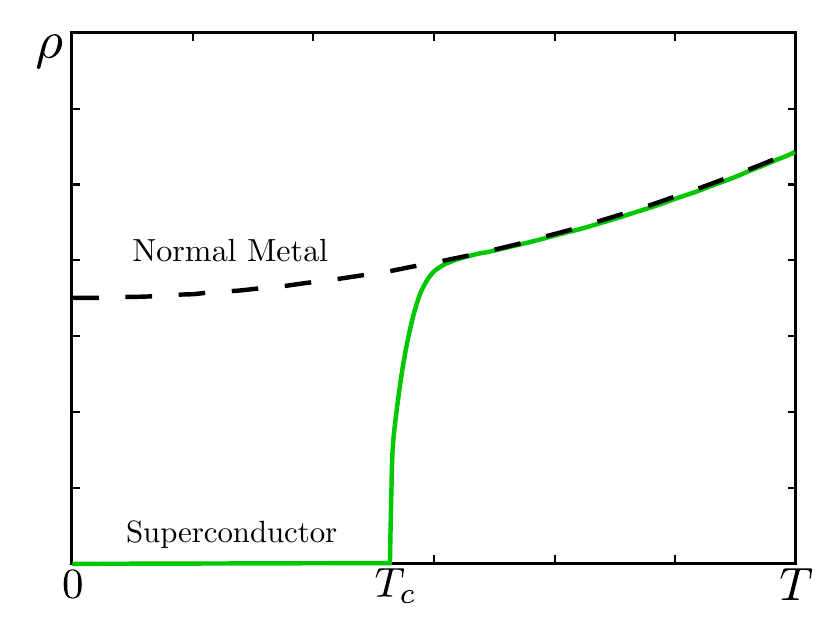}
    \caption{Depiction of a superconducting metal resistivity curve \textit{versus} temperature.}
    \label{fig:hgresistivity}
\end{figure}

The distinction between a perfect conductor and a superconductor is that, while both exhibit zero electrical resistance below $T_c$, their interaction with magnetic fields is phenomenologically different. A perfect conductor would maintain a constant magnetic field in its interior; if cooled in the absence of a field and then exposed to one, it would exclude it, but if cooled in a pre-existing field, it would trap that field internally. A superconductor, in contrast, actively expels all magnetic flux from its interior when cooled below its critical temperature, regardless of the order of operations. This phenomenon of perfect diamagnetism is known as the \textbf{Meissner effect} and is a defining characteristic of superconductivity.

\begin{figure}[!ht]
    \centering
    \includegraphics[width = 76mm]{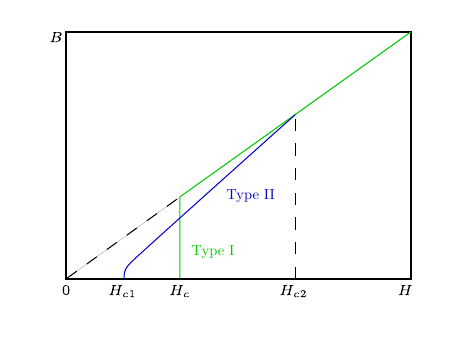}
    \caption{Comparison of the magnetic response for Type I (green) and Type II (blue) superconductors. The plot shows the internal magnetic field ($\mathbf{B}$) versus the applied external field ($\mathbf{H}$). This characteristic behavior assumes a simple geometry, such as a long cylinder with a parallel applied field, to minimize complex effects arising from the sample's shape (demagnetization) and crystalline orientation.}
    \label{fig:suptyp}
\end{figure}

Superconductors can be further defined by their interaction with magnetic flux, being labeled as Type I or Type II. As a qualitative comparison, Type I superconductors completely exclude an applied magnetic flux via the Meissner effect up to a critical field $H_c$, at which point they transition to a normal, resistive state. In contrast, Type II superconductors exhibit two critical fields, $H_{c1}$ and $H_{c2}$. Below $H_{c1}$, they behave like Type I superconductors, completely expelling the magnetic field. However, between $H_{c1}$ and $H_{c2}$, the material enters a "mixed state". In this state, the magnetic field penetrates the material in the form of discrete quantized channels known as \textbf{flux vortices}. Each vortex consists of a normal-state (resistive) core containing magnetic flux, which is surrounded by a circulating supercurrent that shields the rest of the material. This behavior can be visualized in Fig.~\ref{fig:suptyp}.

For applications in superconducting quantum circuits, a hybrid material approach is standard, where the tunneling barrier of a Josephson junction relies on a Type I superconductor, while the surrounding terminals and resonators are frequently fabricated from Type II superconductors to enhance overall device performance and coherence~\cite{Oliver_Welander_2013}. The materials most commonly used for building transmons and other superconducting qubits, such as \textbf{aluminum (Al)} and \textbf{niobium (Nb)}, are Type I or borderline Type I/II~\cite{Koch_2007, kjaergaard_superconducting_2020}. The primary reason is that the qubits' quantum coherence is extremely sensitive to noise~\cite{Wendin_2017}. The “mixed state'' of Type II superconductors, where magnetic flux vortices can penetrate the material, is a significant source of decoherence. The movement of these trapped flux vortices creates magnetic noise which can disrupt the delicate quantum state of the qubit, leading to shorter coherence times~\cite{Sung2019, Galperin_2007}. Therefore, the “cleaner'' magnetic behavior of Type I superconductors, which completely expel magnetic fields, is highly desirable for creating a stable and low-noise environment for the quantum circuit~\cite{Oliver_Welander_2013, kjaergaard_superconducting_2020}. 

The phenomenological description of superconductors has deep implications for its microscopic behavior. In the next section, we explore how the perfect conductivity and the magnetic field expulsion, known as the \textit{Meissner effect}, can be understood in terms of classical physics, and how they require the use of the quantum formalism for a complete description in terms of the London equations.

\begin{table*}[ht!]
\centering
\caption{A selection of common superconducting materials with their respective critical temperatures ($T_c$) and type~\cite{kittel_introduction_2005}.}
\label{tab:superconductors_referenced}
\begin{tabular}{@{}l c c c l@{}}
\toprule
\textbf{Material} & \textbf{Symbol} & \textbf{Critical Temp. ($T_c$)} & \textbf{Type} & \textbf{Notes} \\
\midrule
Aluminum & Al & 1.2 K & I & Widely used for qubits and resonators~\cite{kjaergaard_superconducting_2020}. \\
Niobium & Nb & 9.3 K & II & Borderline type; used for resonators and wiring~\cite{kjaergaard_superconducting_2020}. \\
Tantalum & Ta & 4.5 K & I & Gaining popularity for high-coherence qubits~\cite{place_new_2021}. \\
Lead & Pb & 7.2 K & I & A historically important elemental superconductor. \\
Mercury & Hg & 4.2 K & I & The first superconductor discovered. \\
\addlinespace 
Niobium Nitride & NbN & $\sim$16 K & II & Used in Superconducting Single-Photon Detectors (SNSPDs)~\cite{natarajan_superconducting_2012}. \\
Niobium-Titanium & NbTi & $\sim$10 K & II & The workhorse wire for high-field superconducting magnets~\cite{Patel2019NiobiumtitaniumS}. \\
YBCO & - & $\sim$93 K & II & A common “high-temperature'' superconductor. \\
\bottomrule
\end{tabular}
\end{table*}

\subsection{Classical Formulation}
We'll begin the quantitative development by using a classical formulation from Newtonian Laws and Maxwell equations in order to determine a general behavior pattern for a perfect superconductor.

\subsubsection{Perfect conductivity}

The linear relation between the electric field $\bl{E}$ inside a conductor and the induced current density $\bl{J}$ is quantified by the \textit{conductivity} $\sigma$ according to Ohm's law:
\begin{equation}
    \bl{J} = \sigma \bl{E}.
\end{equation}
At the microscopic level, this relation is understood as the result of the continuous scattering of electrons in the material as described by Drude's model. Perfect conductivity, however, assumes zero-resistivity, frictionless motion of the charge carriers, breaking this relation. From the second law of mechanics, the electrons of \textit{mass} \textit{m} and \textit{charge magnitude} $e>0$ freely accelerate in a field $\bl{E}$ with a rate $\ddot{\bl{r}}$:
\begin{equation}
    \bl{F} = m \ddot{\bl{r}} = - e \bl{E}.
\end{equation}
The \textit{current density} $\bl{J}$ is related to the electron movement and density $n_e$ as:
\begin{equation}
    \bl{J} = - e n_e \dot{\bl{r}} \rightarrow \dot{\bl{J}} = - e n_e \ddot{\bl{r}},
\end{equation}
leading to the relation between $J$ and $E$
\begin{equation}
\label{eq:currentderiv}
    \dot{\bl{J}} = \frac{n_e e^2}{m} \bl{E}.
\end{equation}
This equation is in direct conflict with Ohm's law, since there is no stable value of $\bl{J}$ for an $\bl{E}\neq 0$. 

There are two immediate conclusions: first, since the current density diverges under the action of the electric field, we can understand perfect conductivity as the limit $\sigma\to\infty$. Second, as a consequence, the value of the electric field in the bulk of a perfect conductor must always be null for any finite current flow.

Perfect conductivity also has consequences on the dynamics of the \textit{magnetic field} $\bl{B}$. According to Maxwell equations (Ampère's and Faraday's law, respectively)~\cite{tinkham2004introduction}:
\begin{equation} \label{Eq:maxmagfie}
    \nabla \times \bl{B} = \frac{4\pi}{c} \bl{J},
\end{equation}
\begin{equation} \label{Eq:maxelefie}
    \nabla \times \bl{E} = -\frac{1}{c} \del{\bl{B}}{t},
\end{equation}
where we take the macroscopic average values of the fields, and $c$ is the \textit{speed of light}. From Eqs.  (\ref{eq:currentderiv}) and (\ref{Eq:maxelefie}):
\begin{equation}\label{eq:6}
    \nabla \times \del{\bl{J}}{t}  = - \frac{n_e e^2}{c m} \del{\bl{B}}{t}.
\end{equation}
On the other hand, the time derivative followed by the curl of Eq.(\ref{Eq:maxmagfie}) leads to:
\begin{equation}
    \nabla \times \nabla \times \del{\bl{B}}{t} = \frac{4\pi}{c}\nabla \times \del{\bl{J}}{t},
\end{equation}
which, together with Eq. (\ref{eq:6}), gives:
\begin{equation}
    \nabla \times \nabla \times \del{\bl{B}}{t} = -\frac{4\pi n_e e^2}{m c^2} \del{\bl{B}}{t}.
\end{equation}
This expression can be simplified using the vector identity $\nabla \times \nabla \times=\nabla \nabla \cdot -\nabla^2 $:
\begin{equation}
    \nabla \frac{\partial}{\partial t} (\cancelto{0}{\nabla \cdot \bl{B}}) - \nabla^2 \del{\bl{B}}{t}  = -\frac{4\pi n_e e^2}{m c^2} \del{\bl{B}}{t},
\end{equation}
resulting in the second-order partial differential equation on $\bl{B}$:
\begin{equation} \label{eq:parB}
    \nabla^2 \del{\bl{B}}{t}  = \frac{1}{\lambda^2} \del{\bl{B}}{t},
\end{equation}
where $\lambda$ is a function of the material parameters:
\begin{equation}\label{eq:magpendep}
    \lambda = \sqrt{\frac{m c^2}{4 \pi n_e e^2}}.
\end{equation}
When Eq.\,\ref{eq:magpendep} is analyzed near the surface of the perfect conductor, there are two possible solutions given by $\del{\bl{B}}{t} \propto e^{\pm z/\lambda}$, where $z$ is the spatial coordinate penetrating the material. Neglecting the diverging solution, the conclusion is that the time-derivative of $\bl{B}$ is exponentially suppressed in space with a characteristic penetration depth of $\lambda$, implying the magnetic field is constant in the bulk.

Although this description provides a good insight into the behavior of a perfect conductor, it is an incomplete representation of superconductors, as later experiments would reveal~\cite{Meissner1933, kittel_introduction_2005}. In addition to having zero resistivity, superconductors also show perfect diamagnetism, expelling magnetic flux from their interior. This phenomenon, known as the Meissner effect, is one of the most distinctive properties of a superconductor.

\subsubsection{Meissner effect}

In 1933, Meissner and Oschenfeld observed an exclusion of the magnetic flux when a sphere is cooled below its transition temperature~\cite{Meissner1933, poole2010superconductivity}. This discovery was relevant as it established the fundamental distinction between a superconductor and a hypothetical "perfect conductor". In their experiment, they measured the magnetic field distribution outside single-crystal samples of tin and lead. They observed that when the samples were cooled below their transition temperature while already in an external magnetic field, the magnetic flux was actively expelled from their interior~\cite{Meissner1933}. 

This perfect diamagnetism is a unique characteristic of the superconducting state and is not what would be expected from a perfect conductor, which would merely prevent any change in the internal magnetic field, thus trapping the field if cooled in its presence. The Meissner effect, therefore, revealed that superconductivity is a distinct thermodynamic phase and not simply a limit of classical electromagnetism.

Given the relation between \textit{magnetic flux density} (\bl{H}), \textit{magnetization} (\bl{M}) and \textit{magnetic field} (\bl{B}), a null magnetic field requires, in Gaussian units:
\begin{equation}
    \cancelto{0}{\bl{B}} = \bl{H} + 4\pi \bl{M},
\end{equation}
\begin{equation}\label{eq:meis1}
    \bl{M} = -\frac{1}{4\pi}\bl{H}.
\end{equation}
This relation implies the superconductor currents are rearranged so as to induce a magnetization $\bl{M}$ that completely expels magnetic fields from the bulk of the material. This property, coupled with the expulsion of electric field given by perfect conductivity, has immediate applications in building effective shields to protect sensitive equipment from unwanted electromagnetic interference. Furthermore, the perfect diamagnetism can be utilized in magnetic levitation applications, where objects can float above cooled superconductors, avoiding contact and friction with surfaces. These possibilities pave the way for significant innovations in various fields of technology and science~\cite{sorongane2023classical}. 

Additionally, there's a critical point where \bl{H} exceeds a certain threshold. Beyond this threshold, the material can no longer perfectly cancel the applied magnetic field, leading to the breakdown of superconductivity. In other words, with the increase of $\bl{H}$, after a limit, the material loses its superconductivity and returns to a normal conductive state. This limit can be explained via a quantum formulation in terms of Cooper pairs, which will be discussed in the further section.

\subsection{Quantum Formulation}

In 1951, F. London proposed that the magnetic flux trapped within a superconducting ring is quantized~\cite{london1951}. This quantization is only observed in multiply connected geometries, such as superconducting rings~\cite{deaver1961,doll1961}. While the London equations provide a successful phenomenological description of zero resistance and the Meissner effect, they are ultimately classical and cannot explain all aspects of superconductivity. The first direct evidence of a macroscopic quantum phenomenon in superconductors came from experiments on superconducting rings, which behave identically in this regard regardless of being Type I or Type II. In 1961, two groups independently - B. S. Deaver and W. M. Fairbank~\cite{deaver1961}, and R. Doll and M. Näbauer~\cite{doll1961} - performed a landmark experiment based on a proposal by F. London~\cite{london1951}. They cooled a tiny superconducting ring below its transition temperature while it was threaded by an external magnetic field. When this external field was subsequently removed, a current was induced in the ring to preserve the internal magnetic flux. Because the ring has zero resistance, this current flows indefinitely without dissipation and is known as a \textbf{persistent current}.

The crucial discovery was that the magnetic flux ($\Phi$) trapped by this persistent current was not continuous but was instead quantized, only taking on integer multiples of a fundamental value, the magnetic flux quantum ($\Phi_0 = h/2e$). This observation provided the first clear justification for why a quantum description is necessary. The quantization of a macroscopic variable like magnetic flux is a direct consequence of the underlying quantum mechanics of the superconducting state. The charge carriers (Cooper pairs) are described by a single, macroscopic quantum wave function that possesses a phase. For this wave function to be physically realistic and single-valued, its phase must return to itself plus an integer multiple of $2\pi$ after one complete loop around the ring. This requirement on the phase directly imposes the condition that the magnetic flux contained within the ring must be quantized, linking a macroscopic observable to a quantum mechanical principle.

\subsubsection{Supercurrent Equation}

The quantum description of superconductivity begins with the formulation of a supercurrent, which requires a description of the superconducting charge carriers. The phenomenon arises from the formation of \textbf{Cooper pairs} — bound states of two electrons with opposite momenta and spins — which behave as bosonic quasiparticles~\cite{Bardeen1957}. While the microscopic pairing mechanism may vary, the essential feature is that these pairs, being bosons, condense into a single, coherent macroscopic quantum state.

To phenomenologically describe this collective state, Ginzburg and Landau introduced a \textbf{complex order parameter}, $\Psi(\mathbf{r})$, often called a macroscopic wave function~\cite{Ginzburg1950}. Instead of describing a single particle, $\Psi(\mathbf{r})$ describes the entire condensate of Cooper pairs as a single entity. 

Because $\Psi(\mathbf{r})$ represents a coherent state of quantum particles in the condensate, it obeys a Schr\"odinger-like equation, known as the Ginzburg-Landau equation. In this framework, the charge and mass are not those of a single electron, but the effective charge $q^* = 2e$ and mass $m^* = 2m_e$ of a Cooper pair~\cite{Gorkov1959}. This treatment of a macroscopic system using a single wave function is possible because Bose-Einstein statistics allow all Cooper pairs to occupy the same quantum ground state, enforcing the system-wide phase coherence that is characteristic of superconductivity~\cite{tinkham2004introduction}.

Delving deeper into this formalism, the order parameter has its squared modulus proportional to the local Cooper pair density. Since each Cooper pair consists of two fermions, the density of Cooper pairs $n_s = N/V$, where $N$ is the total number of Cooper pairs and $V$ is the volume of the superconductor, is half the density of superconducting electrons. Therefore, the squared modulus of the order parameter satisfies~\cite{Ginzburg1950}:
\begin{equation}\label{eq:psi_dens}
    |\Psi (\bl{r})|^2 = n_s.
\end{equation}
The current associated with the superconducting state arises from the coupling of the Cooper pair condensate to the electromagnetic field. In the presence of a magnetic vector potential $\bl{A}$, the canonical momentum operator acting on the condensate is given by:
\begin{equation}
    \hat{\mathbf{p}}_M = -i\hbar\boldsymbol{\nabla} - \frac{2e}{c} \mathbf{A},
\end{equation}
where $2e$ is the charge of a Cooper pair. The supercurrent density $\bl{J}$ is then expressed as:
\begin{eqnarray}
\bl{J} &=& \frac{2e}{2m} \left[ \Psi^* \left( \frac{\hbar}{i} \nabla - \frac{2e}{c} \bl{A} \right) \Psi 
+ \mathrm{h.c.} \right],
\label{Eq:supcurdef}
\end{eqnarray}
where $2m$ is the effective mass of the Cooper pair, and “h.c.” denotes the Hermitian conjugate.

Under conditions where the spatial variation of the condensate density is negligible — i.e., in the absence of significant external perturbations or near the center of the superconducting phase — the order parameter can be approximated as:
\begin{equation} \label{Eq:supcurwav}
    \Psi (\bl{r}) \approx \sqrt{n_s} e^{i \theta(\bl{r})}, 
\end{equation}
where $\theta(\bl{r})$ is the local phase of the condensate. This macroscopic phase, $\theta(\mathbf{r})$, is the single most important degree of freedom for describing superconducting quantum circuits. As will be detailed in Sec. \ref{sec:josephson}, a difference in this phase across a tunnel junction gives rise to the \textbf{Josephson effect}, which is the fundamental principle enabling the creation of superconducting qubits. Substituting Eq.~(\ref{Eq:supcurwav}) into Eq.~(\ref{Eq:supcurdef}) yields the supercurrent density:
\begin{equation} \label{eq: supcurr}
    \bl{J} = -\frac{\bl{A}}{\Lambda_s c} + \frac{\hbar}{2e \Lambda_s}\nabla\theta,
\end{equation}
\begin{equation} \label{eq:penetrationDepth}
    \Lambda_s = \frac{4\pi \lambda^2}{c^2} = \frac{2m}{n_s (2e)^2} = \frac{m}{2 n_s e^2},
\end{equation}
where $\lambda$ is the London penetration depth, and the factors $2e$ and $2m$ explicitly account for the charge and mass of Cooper pairs, respectively. This expression connects the supercurrent to the phase of the order parameter and explains the phenomenon of flux quantization: Integrating Eq.~\ref{eq: supcurr} over a closed loop inside the ring material yields $\oint \nabla\theta \cdot d\mathbf{l} = (2e/\hbar) \oint \mathbf{A} \cdot d\mathbf{l}$. Since the wave function must be single-valued, the total change in phase around the loop must be an integer multiple of $2\pi$. This forces the magnetic flux $\Phi = \oint \mathbf{A} \cdot d\mathbf{l}$ to be quantized in units of the flux quantum, $\Phi_0 = h/2e$.

The uniformity of $n_s$ is justified under equilibrium conditions well below the critical temperature and in the absence of strong spatial inhomogeneities. In this regime, the number of condensed pairs remains approximately constant throughout the material, leading to a constant Cooper pair density $n_s$. 

When an external magnetic field is applied to a superconductor, this formalism explains the Meissner effect. A persistent supercurrent, described by Eq.~\ref{eq: supcurr}, is induced at the surface of the material. This \textbf{screening current} flows such that it generates a magnetic field that exactly cancels the external field inside the superconductor, ensuring the magnetic field in the bulk remains zero. This interaction is central to the diamagnetic behavior of superconductors~\cite{tinkham2004introduction, poole2010superconductivity}.

\subsubsection{London Equations}

From the quantum mechanical formulation of the supercurrent, both characteristic effects of superconductors — perfect conductivity and perfect diamagnetism — can be derived, leading to the so-called London equations.

To derive the first London equation, consider a time-independent Cooper pair density and a perturbation introduced by an external scalar electric potential $\phi(\bl{r})$. Under these conditions, the Schrödinger equation for the superconducting condensate, whose wave function is denoted by $\Psi(\bl{r}, t) = \sqrt{n_s} e^{i\theta(\bl{r}, t)}$, can be written as:
\begin{equation}\label{eq:ham_perturb}
    \left[ \hat{H}_0 + e \phi(\mathbf{r}) \right] \Psi(\mathbf{r}, t) = i\hbar \frac{\partial}{\partial t} \Psi(\mathbf{r}, t),
\end{equation}
where $\hat{H}_0$ denotes the unperturbed Hamiltonian of the system, and $e\phi(\bl{r})$ represents the perturbation due to an applied electric potential. Substituting the wave function into Eq.~(\ref{eq:ham_perturb}) and isolating the temporal evolution of the phase leads to:
\begin{equation}
    \hbar \frac{\partial \theta(\mathbf{r}, t)}{\partial t} = - \left[ E_0 + e \phi(\mathbf{r}) \right],
\end{equation}
where \(E_0\) is the unperturbed energy eigenvalue, independent of position.

Differentiating the supercurrent expression (Eq.~\ref{eq: supcurr} with respect to time, yields:
\begin{equation}
    \frac{\partial \bl{J}}{\partial t} = -\frac{1}{\Lambda_s c} \frac{\partial \bl{A}}{\partial t} + \frac{\hbar}{e \Lambda_s} \nabla \frac{\partial \theta}{\partial t}.
\end{equation}
Using the phase evolution relation:
\begin{equation}
    \frac{\partial \bl{J}}{\partial t} = -\frac{1}{\Lambda_s c} \frac{\partial \bl{A}}{\partial t} - \frac{1}{\Lambda_s} \nabla \phi(\bl{r}).
\end{equation}

Recognizing the electric field as:
\begin{equation}
    \bl{E} = -\frac{1}{c} \frac{\partial \bl{A}}{\partial t} - \nabla \phi(\bl{r}),
\end{equation}
the relation becomes:
\begin{equation}\label{eq:london1}
    \bl{E} = \Lambda_s \frac{\partial \bl{J}}{\partial t},
\end{equation}
which constitutes the \textbf{first London equation}, capturing the ideal conductivity of superconductors. This equation implies that an electric field induces a proportional change in the supercurrent density, with the proportionality governed by $\Lambda_s$, a material-specific parameter.

Two main implications follow:
\begin{itemize}
    \item An applied electric field leads to a continuous acceleration of the supercurrent, in stark contrast to normal resistive behavior. This illustrates the phenomenon of zero electrical resistance in superconductors.
    \item The evolution of the supercurrent is directly determined by the local electric field, establishing a dynamic response rather than a dissipative one.
\end{itemize}

To derive the \textbf{second London equation}, which characterizes perfect diamagnetism, consider the curl of the supercurrent expression:
\begin{equation}
    \nabla \times \bl{J} = -\frac{1}{\Lambda_s c} \nabla \times \bl{A} + \frac{\hbar}{e \Lambda_s} \nabla \times \nabla \theta.
\end{equation}
Since the curl of a gradient vanishes,
\begin{equation}
    \nabla \times \bl{J} = -\frac{1}{\Lambda_s c} \bl{B}.
\end{equation}

Taking the curl of both sides and using Ampère's Law, Eq.~(\ref{Eq:maxmagfie}), it yields:
\begin{equation}
    \nabla \times \left( \frac{c}{4\pi} \nabla \times \bl{B} \right) = -\frac{1}{\Lambda_s c} \bl{B}.
\end{equation}

Using the vector identity $\nabla \times (\nabla \times \bl{B}) = \nabla(\nabla \cdot \bl{B}) - \nabla^2 \bl{B}$ and Gauss' law for magnetism $\nabla \cdot \bl{B} = 0$, the expression reduces to:
\begin{equation}
    -\frac{c}{4\pi} \nabla^2 \bl{B} = -\frac{1}{\Lambda_s c} \bl{B}.
\end{equation}

Rearranging, and defining the \textit{London penetration depth} $\lambda$ as in Eq.~(\ref{eq:penetrationDepth}), one obtains:
\begin{equation}\label{eq:london2}
    \nabla^2 \bl{B} = \frac{1}{\lambda^2} \bl{B}.
\end{equation}

The second London equation describes how magnetic fields decay exponentially inside a superconductor, a direct manifestation of the Meissner effect. This result confirms that superconductors exhibit perfect diamagnetism by completely expelling magnetic fields from their interior (except within a thin surface layer of thickness $\lambda$).

The second London equation, $\nabla^2 \mathbf{B} = \mathbf{B}/\lambda^2$, has a straightforward physical interpretation. For a simple geometry, such as a magnetic field parallel to the surface of a semi-infinite superconductor, its solution shows that the field decays exponentially from the surface into the bulk: $B(x) = B(0)e^{-x/\lambda}$. The parameter $\lambda$ is therefore the characteristic length scale over which an external magnetic field is screened. This depth is a crucial material property. In conventional superconductors used for quantum circuits, like aluminum and niobium, $\lambda$ is typically in the range of 50-100 nm~\cite{poole2010superconductivity}. For high-temperature cuprate superconductors such as YBa$_2$Cu$_3$O$_{7-\delta}$, the penetration depth can be larger, often exceeding 200 nm~\cite{Basov2005}.

This short-scale magnetic screening is fundamental to modern quantum technologies. It allows for the fabrication of ultra-thin superconducting films and devices where the bulk of the material remains perfectly diamagnetic and shielded from external fields. This property is heavily exploited in the design of superconducting qubits and resonators, ensuring a low-noise, high-coherence environment while still allowing for strong electromagnetic coupling in confined geometries~\cite{Devoret2013}.

Together, the London equations—derived from the quantum treatment of the superconducting condensate—encapsulate the two defining properties of superconductors: zero electrical resistance and perfect diamagnetism. They describe how a superconductor responds to external electric and magnetic fields, providing the foundational framework for understanding its electromagnetic behavior.

Together, the London equations — derived from the quantum treatment of the superconducting condensate — encapsulate the two defining properties of superconductors: zero electrical resistance and perfect diamagnetism. They describe how a superconductor responds to external magnetic fields by generating surface currents that exactly cancel the interior field, leading to the magnetic field's expulsion.

\section{Josephson Effect} \label{sec:josephson}
The Josephson effect, discovered by physicist Brian D. Josephson~\cite{JOSEPHSON1962251} in the early 1960s, involves the phase-dependent behavior of electric currents in superconductors. Afterward, Philip W. Anderson recognized the significance of Josephson's work, exploring the theoretical concepts behind the effect~\cite{anderson1963,ANDERSON19671}. This discovery has fundamental implications for quantum coherence in macroscopic systems and has practical applications in fields such as metrology and superconducting electronics.

\begin{figure}[!h]
    \centering
    \includegraphics[width = 76mm]{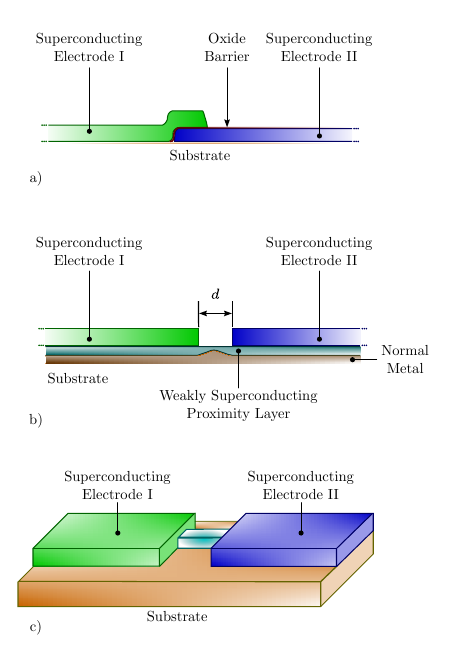}
    \caption{Three types of Josephson junction: a) Superconductor-Insulator-Superconductor; b) Superconductor-Normal-Superconductor; c) Superconductor-Constriction-Superconductor~\cite{tinkham2004introduction}.}
    \label{fig:jjexamples}
\end{figure}

The Josephson effect arises from the quantum mechanical tunneling of Cooper pairs across a weak coupling interface — termed a \textit{Josephson junction} — connecting two bulk superconductors~\cite{timm2023theory}. A Josephson junction is fundamentally characterized by a non-superconducting region (the “weak link”) that permits phase-coherent coupling of the superconducting order parameters. This weak link may manifest as: (i) an insulating barrier, enabling tunneling through a classically forbidden region; (ii) a normal metal layer, rendered superconducting via the proximity effect through Andreev reflection processes; or (iii) a geometrically constrained region (e.g., a nanoscale constriction) within an otherwise homogeneous superconducting medium~\cite{tinkham2004introduction}. Examples of these configurations can be visualized in Fig.~\ref{fig:jjexamples}.

The junction’s defining property—its critical current-phase relation—emerges from the interplay between gauge symmetry breaking and macroscopic quantum coherence, establishing it as a cornerstone of superconducting quantum electronics.

\subsection{DC Josephson Effect}

A basic quantitative formulation of the Josephson effect begins with the assumption of two superconductors, labeled A and B, weakly coupled through a thin insulating barrier C. This configuration results in a potential barrier that is classically forbidden for Cooper pairs, making their transfer across the junction possible only via quantum tunneling.

Assuming that both superconductors A and B have well-defined numbers of Cooper pairs, denoted by $N_A$ and $N_B$ respectively, each superconductor can be described by its corresponding Fock state $|N_A\rangle$ and $|N_B\rangle$. These are eigenstates of the respective number operators, and also, under suitable conditions, eigenstates of the unperturbed Hamiltonians of each isolated superconductor. The total quantum state of the coupled system is then given by:
\begin{equation}
    |\psi\rangle = |N_A\rangle \otimes |N_B\rangle \equiv |N_A, N_B\rangle.
\end{equation}

If the tunneling of $m$ Cooper pairs from A to B is considered (such that the total number of particles remains conserved), the resulting state becomes:
\begin{equation}
    |m\rangle = |N_A - m, N_B + m\rangle.
\end{equation}
This framework models tunneling as a coherent process that changes the particle number in each superconductor symmetrically. States of the form $|m\rangle$ form a complete basis for the Hilbert space of the junction.

The tunneling process can then be described by a Hamiltonian of the form:
\begin{equation}\label{eq:jjmham}
    \hat{H}_T = -\frac{1}{2} E_J \sum_m \left[|m\rangle \langle m+1| + |m+1\rangle \langle m|\right],
\end{equation}
where $E_J$ is the Josephson energy, associated with the tunneling of a single Cooper pair across the junction. Higher-order terms of the form $|m\rangle \langle m+n|$ with $n > 1$ are neglected under the assumption that the tunneling is weak, making multiple simultaneous pair transitions highly improbable and their contributions exponentially suppressed. This justifies a nearest-neighbor coupling model.

To diagonalize $\hat{H}_T$, consider its analogy with a one-dimensional tight-binding Hamiltonian with hopping amplitude $-E_J/2$ and lattice spacing $1$. The eigenstates of such a Hamiltonian are delocalized Bloch-like states defined by:
\begin{equation}
    |\theta\rangle = \sum_{m = -\infty}^{\infty} e^{i m \theta} |m\rangle,
\end{equation}
where $\theta \in [0, 2\pi)$ is the conjugate variable to $m$, interpreted as a phase. The inverse transformation is given by:
\begin{equation} \label{eq:invm}
    |m\rangle = \frac{1}{2\pi} \int_0^{2\pi} e^{-i m \theta} |\theta\rangle \, d\theta.
\end{equation}

In this basis, the tunneling Hamiltonian becomes diagonal:
\begin{equation}
    \hat{H}_T = - E_J \cos \hat{\theta},
\end{equation}
where the operator $e^{i \hat{\theta}}$ is defined by:
\begin{equation} \label{eq:eioop}
    e^{i \hat{\theta}} = \frac{1}{2 \pi} \int_0^{2\pi} e^{i \theta} |\theta\rangle \langle \theta| \, d\theta.
\end{equation}

At this point, it is essential to provide a physical interpretation of the phase variable $\theta$. While the $\theta$ appearing in the phase basis $|\theta\rangle$ is a mathematical construct arising from the Fourier transform of number states $|m\rangle$, it gains physical relevance when related to the macroscopic quantum phase difference across the Josephson junction. This identification becomes meaningful when the superconducting phase is introduced via the Ginzburg-Landau theory.

\subsubsection*{Connection to Superconducting Phase and Magnetic Flux}

Recalling the macroscopic phase $\theta$ of the order parameter introduced in Eq.~(\ref{Eq:supcurwav}), the essential physics of a Josephson junction emerges when a weak link connects two superconductors. When two superconductors $A$ and $B$ are connected via a weak link, the difference in their macroscopic phases, $\varphi = \theta_A - \theta_B$, acquires physical significance and determines the behavior of the junction, where $\theta_A$ and $\theta_B$ are the respective macroscopic phases of each superconductor.

Given this perspective, the \textit{reduced magnetic flux} $\varphi$, the \textit{magnetic flux} $\phi$, and the \textit{reduced flux quantum} $\phi_0$ are defined by:
\begin{equation}
    \varphi = \frac{\phi}{\phi_0}, \qquad \phi_0 = \frac{\hbar}{2e}.
\end{equation}
The reduced flux $\varphi$ is directly related to the phase difference across the junction, and one identifies:
\begin{equation}
    \theta \equiv \varphi \mod{2\pi}, \qquad \varphi = \theta_A - \theta_B,
\end{equation}
where $\theta_A$ and $\theta_B$ are the Ginzburg-Landau phases associated with superconductors A and B, respectively.

This identification justifies interpreting the operator $\hat{\theta}$ as representing the phase difference across the junction and establishes $\varphi$ as its physically relevant counterpart. Therefore, the tunneling Hamiltonian can be rewritten in terms of the reduced flux:
\begin{equation}\label{eq:jjham}
    \hat{H}_T = - E_J \cos \hat{\varphi}.
\end{equation}
where $\hat{\varphi}$ denotes the quantum operator associated with the gauge-invariant phase difference across the junction. Within the canonical quantization framework, $\hat{\varphi}$ and the number operator $\hat{n}$, which accounts for the net Cooper pair transfer across the junction, satisfy the commutation relation~\cite{kjaergaard_superconducting_2020}
\begin{equation}\label{eq:varphiCommutation}
    [\hat{\varphi}, \hat{n}] = i,
\end{equation}
analogous to the canonical position-momentum relation in quantum mechanics. This conjugate structure imposes a fundamental quantum uncertainty between phase and charge, precluding their simultaneous precise determination.

\subsubsection*{Derivation of the DC Josephson Relation}

To derive the expression for the supercurrent, consider a wave packet in the phase basis, written as:
\begin{equation}
    |\psi\rangle = \sum_{\varphi} \mu_{\varphi} |\varphi\rangle,
\end{equation}
where $\mu_{\varphi}$ are complex amplitudes over the continuous phase variable $\varphi$. The group velocity of this wave packet is given by the derivative of the energy expectation value with respect to the conjugate variable $\varphi$. Accordingly, the expectation value of the current operator $I_s$ takes the form
\begin{equation}
    I_s = 2e v_g = \frac{2e}{\hbar} \frac{\partial}{\partial \varphi} \langle \hat{H}_T \rangle.
\end{equation}
Substituting the tunneling Hamiltonian from Eq.~(\ref{eq:jjham}), one obtains the \textbf{DC Josephson relation}:
\begin{equation}\label{eq:dcjos}
    I_s = I_c \sin \varphi,
\end{equation}
where $\varphi = \langle \hat{\varphi} \rangle$ denotes the expected value of the phase difference operator across the junction. The critical current $I_c$ is defined as
\begin{equation}
    I_c = \frac{2e E_J}{\hbar}.
\end{equation}

\subsubsection*{Physical Interpretation}

The DC Josephson relation, Eq.~(\ref{eq:dcjos}), encapsulates two key features of coherent charge transport in Josephson junctions:
\begin{itemize}
    \item A non-zero supercurrent can flow across the junction even in the absence of an applied voltage, provided there exists a finite phase difference $\varphi$ between the superconducting leads;
    \item The maximum achievable supercurrent is given by the critical current $I_c$, beyond which the junction enters a dissipative regime.
\end{itemize}

Importantly, Eq.~(\ref{eq:dcjos}) is valid in the semiclassical regime where the phase difference is well-defined and quantum fluctuations are negligible. It does \emph{not} require the system to be in an eigenstate of the tunneling Hamiltonian $\hat{H}_T$. In fact, the energy eigenstates of $\hat{H}_T$ are delocalized in phase space and exhibit a vanishing expectation value of the current operator due to symmetry. The DC Josephson relation describes the current in coherent or semiclassical states, characterized by a narrow distribution in phase.

Moreover, in time-dependent scenarios where the phase $\varphi$ evolves according to:
\begin{equation}
    \frac{d\varphi}{dt} = \frac{2e}{\hbar} V(t),
\end{equation}
the system is no longer in a stationary state. Nevertheless, the instantaneous current still obeys Eq.~(\ref{eq:dcjos}) under the assumption that the time evolution remains adiabatic with respect to the junction dynamics.

The Josephson effect thus exemplifies macroscopic quantum coherence, manifesting as a non-dissipative current determined solely by the phase difference. This phenomenon underpins the operation of superconducting quantum circuits and serves as a cornerstone in the architecture of superconducting qubits.

\subsection{AC Josephson Effect}

The application of a static voltage \( V \) across a Josephson junction induces coherent temporal dynamics in the superconducting phase difference. This behavior arises from the fundamental conjugate relation between the superconducting phase operator \(\hat{\varphi}\) and the number operator \(\hat{n}\), given by Eq. \ref{eq:varphiCommutation}. When a constant voltage is applied, the Hamiltonian describing the junction becomes:
\begin{equation}
    \hat{H} = -E_J \cos \hat{\varphi} - 2eV \hat{n},
\end{equation}
where the second term accounts for the potential energy associated with the transfer of Cooper pairs across the voltage bias.

To determine the resulting phase dynamics, consider the Heisenberg equation of motion for \(\hat{\varphi}(t)\):
\begin{equation}
    \frac{d\hat{\varphi}}{dt} = \frac{i}{\hbar} [\hat{H}, \hat{\varphi}] = \frac{2e}{\hbar} V.
\end{equation}
This yields a linear time evolution for the operator:
\begin{equation}
    \hat{\varphi}(t) = \hat{\varphi}_0 + \frac{2eV}{\hbar} t.
\end{equation}

To obtain a measurable quantity, one considers the expectation value of the phase operator in a semiclassical state where quantum fluctuations are negligible. Denoting the expectation value by \(\varphi(t) = \langle \hat{\varphi}(t) \rangle\), it follows that:
\begin{equation} \label{eq:acjos}
    \frac{d\varphi}{dt} = \frac{2e}{\hbar} V,
\end{equation}
which defines the \textbf{AC Josephson relation}.

Integrating Eq.~\eqref{eq:acjos}, the phase difference evolves as:
\begin{equation}
    \varphi(t) = \varphi_0 + \omega_J t,
\end{equation}
where \(\omega_J = 2eV/\hbar\) is the \textit{Josephson frequency} associated with the applied voltage.

Substituting this time-dependent phase into the first Josephson relation yields the supercurrent:
\begin{equation}
    I_s(t) = I_c \sin \varphi(t) = I_c \sin(\varphi_0 + \omega_J t).
\end{equation}

The following conclusions emerge from this analysis:
\begin{itemize}
    \item A constant voltage bias across the junction results in an alternating supercurrent of amplitude \( I_c \) and frequency \( \omega_J \);
    \item The Josephson junction functions as a voltage-to-frequency converter, with conversion factor \( f_J = (2e/h)V \);
    \item The AC Josephson effect forms the physical basis for high-precision voltage standards in quantum metrology.
\end{itemize}

This phenomenon exemplifies macroscopic quantum coherence, linking phase dynamics and electrical transport in a fundamentally quantum mechanical manner. It also underpins the operation of superconducting devices at microwave and radio frequencies, enabling technologies ranging from quantum voltage standards to superconducting qubits.

\section{Circuit Quantum Electrodynamics} \label{sec:CQE}
The preceding sections have comprehensively covered the foundational aspects of superconductivity, ranging from classical formulations to the quantum phenomena known as the Josephson effect. Building upon this groundwork, the forthcoming segment delves into Circuit Quantum Electrodynamics (cQED)~\cite{Blais_2021}, marking a transition from theoretical underpinnings to practical applications in the realm of electrical circuits and quantum mechanics. In this section, the focus shifts towards an in-depth exploration of different devices within the context of cQED such as resonators and transmons. 

\subsection{Resonator}
Resonators, pivotal components in quantum circuits, exhibit intricate interactions with microwave photons reminiscent of those between atoms and light in cavity quantum electrodynamics~\cite{Haroche2020}. The upcoming section initially navigates through resonators classically, likening their behavior to classical oscillatory systems and electromagnetic waves. Transitioning to a quantum perspective, it delves into the interaction of quantized electromagnetic fields with matter, elucidating resonators' unique quantum phenomena. This exploration extends to resonators' integration into transmission line architectures, elucidating their pivotal role in coherent quantum information manipulation and transmission.

\subsubsection{Classical Resonator}

\begin{figure}[!ht]
    \centering
    \includegraphics[width = 76mm]{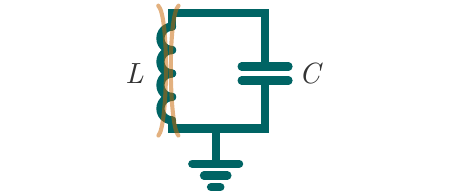}
    \caption{Schematic representation of a classical LC resonator. The inductance $L$ and capacitance $C$ are arranged in parallel, forming an idealized resonant circuit.}
    \label{fig:clasres}
\end{figure}

To understand the classical behavior of a resonator, consider an LC circuit composed of an inductor with inductance $L$ and a capacitor with capacitance $C$ arranged in parallel. This configuration is illustrated in Fig.~\ref{fig:clasres}.

Assign a time-dependent voltage $V(t)$ and flux $\phi(t)$ to the top node of the resonator, with a current $I(t)$ flowing in the counter-clockwise direction. In the absence of dissipation, the total energy in the system is stored alternately in the electric field of the capacitor and the magnetic field of the inductor.

Using Kirchhoff’s laws, the dynamics of the LC circuit can be described by a second-order differential equation. This is the equation of a harmonic oscillator, whose natural angular frequency is given by:
\begin{equation}
    \omega_r = \frac{1}{\sqrt{LC}}.
\end{equation}

The general solutions for the time evolution of current and voltage in the circuit are:
\begin{equation}
    I(t) = I_0 \cos(\omega_r t + \vartheta),
\end{equation}
\begin{equation}
    V(t) = L \frac{dI}{dt} = -L \omega_r I_0 \sin(\omega_r t + \vartheta),
\end{equation}
where $I_0$ is the amplitude of the oscillating current, $V_0 = L \omega_r I_0$ is the amplitude of the voltage, and $\vartheta$ is the initial phase, determined by the initial conditions of the system.

The \textit{characteristic impedance} of the resonator is defined as the ratio between the voltage and current amplitudes:
\begin{equation}\label{eq:resonatorImpedance}
    Z_r = \frac{V_0}{I_0} = \sqrt{\frac{L}{C}}.
\end{equation}

This impedance plays a central role in the coupling of the resonator to external circuits and to quantum systems, as will be discussed in subsequent sections.

In addition, the quality factor $Q$, which is the ratio of its resonant frequency $\omega_r$ to its decay rate $\kappa$, quantifies the resonator’s energy retention capacity by comparing the time-averaged stored energy to the energy dissipated per oscillation cycle. In practical systems, dissipation arises from resistive losses, introducing a decay rate. This modifies the ideal harmonic oscillator equation to a damped form, governing the exponential decay of oscillations. For a parallel $RLC$ configuration, $Q$ is directly proportional to the resistance $R$ and inversely proportional to the characteristic impedance $Z_r$.

The idealized lossless resonator ($R \to \infty$) corresponds to $Q \to \infty$, reflecting perpetual energy exchange. In quantum-electrodynamic applications, high $Q$-values ($Q \gg 1$) are critical for minimizing photon loss and decoherence, as they ensure long-lived electromagnetic field confinement.

This classical resonator model provides the foundation for understanding superconducting resonators and their role in the cQED architecture used in superconducting qubits.

\subsubsection{Quantum Resonator – Simple Formulation}
The quantum formulation of the resonator highlights the transition from classical to quantum descriptions, emphasizing the scales at which quantum effects become significant. While classical physics dominates at macroscopic scales, the focus shifts to systems where quantum mechanics reveals its granularity. By quantizing flux and charge variables through employing second quantization techniques, it is possible to construct the Hamiltonian operator to accurately capture the resonator's quantum behavior. Alas, the exploration of zero-point energy and Gaussian probability distributions further elucidates the quantum nature of the system, providing insights into fundamental phenomena.

Within this perspective, a quantum formulation is built to approach the resonator. We start with the \textit{classical Hamiltonian} $H$ of the parallel LC circuit:
\begin{equation} 
    H = \frac{Q^2}{2C} + \frac{1}{2} C \omega_r^2\phi^2.
\end{equation}

In this formulation, the classical Hamiltonian $H$ for the parallel LC circuit mirrors the form of the Hamiltonian for a harmonic oscillator, which is depicted in Appendix \ref{secA1}, Eq. (\ref{eq:class_harm_osc}). The parallel between the two systems becomes evident when comparing their respective energy expressions. In the case of the LC circuit, $H$ consists of the sum of kinetic and potential energy terms, akin to the expression for a harmonic oscillator. Specifically, the first term corresponds to the kinetic energy term, while the second term represents the potential energy term. This parallel underscores the resonator's behavior akin to that of a harmonic oscillator, albeit in the context of electrical circuits.

Using the Hamilton's equations of motion, it is possible to verify that the flux and the charge variables constitute a pair of canonical conjugated variables, where they play a role of generalized position and generalized momentum, respectively. This verification is seen below:
\begin{equation}
    \del{H}{\phi} = C \omega_r^2 \phi = i = - \frac{dQ}{dt},
\end{equation}
\begin{equation}
    \del{H}{Q} = \frac{Q}{C} = V = \frac{d\phi}{dt}.
\end{equation}

The \textit{flux operator} $\hat{\phi}$ and the \textit{charge operator} $\hat{Q}$ are defined so that they obey the canonical commutation relation:
\begin{equation}
    [\hat{\phi}, \hat{Q}] = i \hbar, 
\end{equation}
and we define the dimensionless operators $\Tilde{\phi}$ and $\Tilde{Q}$:
\begin{equation}
    \Tilde{\phi} = \sqrt{\frac{C \omega_r}{2\hbar}} \hat{\phi},
\end{equation}
\begin{equation}
    \Tilde{Q} = \sqrt{\frac{1}{2 \hbar C \omega_r}} \hat{Q},
\end{equation}
with the commutation relation:
\begin{equation}
    [\Tilde{\phi}, \Tilde{Q}] = \frac{i}{2}. 
\end{equation}

Given that, the Hamiltonian for the resonator becomes:
\begin{equation} \label{eq:resham}
    \hat{H} = \hbar \omega_r (\Tilde{Q}^2 + \Tilde{\phi}^2),
\end{equation}
manipulating Eq. (\ref{eq:resham}), it can be rewritten as:
\begin{equation}\label{eq:reshamfac}
    \hat{H} = \hbar \omega_r \left[ (\Tilde{\phi} - i \Tilde{Q})(\Tilde{\phi} + i \Tilde{Q}) + \frac{1}{2}  \right].
\end{equation}

Relating the Eq. (\ref{eq:reshamfac}) with the \textit{annihilation and creation operators} $\hat{a}$ and $\hat{a}^\dagger$, respectively, the following relations can be derived:
\begin{equation}
   \left \{  \begin{array}{cc}
        \hat{a} &= \Tilde{\phi} + i\Tilde{Q}  \\
         \hat{a}^\dagger &= \Tilde{\phi} - i\Tilde{Q} 
    \end{array}
    \right.,
\end{equation}
\begin{equation}\label{eq:phiop}
    \hat{\phi} = \sqrt{\frac{\hbar}{2 C \omega_r}}(\hat{a}^\dagger + \hat{a}),
\end{equation}
\begin{equation}\label{eq:qop}
    \hat{Q} = i \sqrt{\frac{\hbar C \omega_r}{2}}(\hat{a}^\dagger - \hat{a}),
\end{equation}
\begin{equation}
    \hat{H} = \hbar \omega_r \left(\hat{a}^\dagger \hat{a} + \frac{1}{2} \right),
\end{equation}
where the commutation relation between $\hat{a}^\dagger$ and $\hat{a}$ can be seen in Eq. (\ref{eq:number_commutation}).

Additionally, some useful relations can also be derived within the zero point energy, where $\phi_{zpf}$ represents the \textit{zero-point fluctuation of the flux variable} in the resonator, and $Q_{zpf}$ represents the \textit{zero-point fluctuation of the charge variable} in the resonator. These terms refer to the values that characterize the minimum fluctuation or uncertainty in the flux and charge variable due to the Heisenberg uncertainty principle, respectively:
\begin{equation}
    \big\langle\Tilde{\phi}\big\rangle = \big\langle\Tilde{Q}\big\rangle = 0,
\end{equation}
\begin{equation}
    \big\langle\Tilde{\phi}^2\big\rangle = \phi_{zpf}^2 = \frac{\hbar}{2 C \omega_r},
\end{equation}
\begin{equation}
    \big\langle\Tilde{Q}^2\big\rangle = Q_{zpf}^2 = \frac{\hbar C \omega_r}{2}.
\end{equation}

Moreover, it is conceived an analogy to  the mechanical variables in a Harmonic Quantum Oscillator. Defining $|\phi \ket$ and $|q \ket$ eigenstates of flux and charge:
\begin{equation}
    \bra \phi' | \hat{\phi} | \phi \ket = \phi \delta(\phi' - \phi),  
\end{equation} 
\begin{equation}
    \bra \phi' | \hat{Q} | \phi \ket = -i\hbar \delta(\phi' - \phi) \nabla.
\end{equation}

Projecting the zero point energy eigenstates on the respective eigenstates:
\begin{equation}\label{eq:quares01}
    \bra \phi | 0 \ket = \frac{1}{\phi_{zpf}}\frac{1}{\sqrt{2\pi}} \exp{\left(-\frac{\phi^2}{2\phi_{zpf}^2}\right)},
\end{equation}
\begin{equation}\label{eq:quares02}
    \bra q | 0 \ket = \frac{1}{Q_{zpf}}\frac{1}{\sqrt{2\pi}} \exp{\left(-\frac{q^2}{2Q_{zpf}^2}\right)}.
\end{equation}

Following the quantum formulation built until now, the understanding of what is happening with the flux in the inductor and the charge in the capacitor in every number state is feasible. In a coherent state $|z\ket$, the actual oscillation is observed. This observation stems from the nature of coherent states in quantum mechanics (as described in Appendix \ref{secA1}).

Coherent states are defined as the eigenstates of the non-Hermitian annihilation operator $\hat{a}$, satisfying $\hat{a} |z\ket = z |z\ket$. They mimic the classical oscillations with a well-defined phase and amplitude. Thus, this state has a null average value when you measure the distribution of the quadrature operators, which represent the canonical conjugate observables associated with phase space, and the standard deviation in the ground state is determined by the zero-point fluctuation values:
\begin{equation}
    |z\ket = e^{-\frac{|z|^2}{2}} \sum_{n=0}^\infty \frac{z^n}{\sqrt{n!}} |n\ket.
\end{equation}

The time evolution of such a state is so that its \textit{eigenvalue} $z$ gains a phase factor:
\begin{equation}
    |z(t)\ket = e^{-i\omega_r t} |z(0)\ket,
\end{equation}
where the average values of flux and charge are given as functions of time by:\begin{subequations}
\begin{equation}
 \big\langle\phi(t)\big\rangle = \phi_{zpf} \bra z(t)| (\hat{a}^\dagger + \hat{a}) |z(t)\ket
= 2\phi_{zpf} \mathfrak{Re}\left\{e^{-i\omega_r t} z(0)\right\},
\end{equation}
\begin{equation}
    \big\langle\phi(t)\big\rangle = 2\phi_{zpf} |z(0)| \cos(\omega_r t + \xi),
\end{equation}
\begin{equation}
\big\langle Q(t)\big\rangle = iQ_{zpf} \bra z(t)| (\hat{a}^\dagger - \hat{a}) |z(t)\ket
= 2Q_{zpf} \mathfrak{Im}\left\{e^{-i\omega rt} z(0)\right\},
\end{equation}
\begin{equation}   
   \big\langle Q(t)\big\rangle = 2Q_{zpf} |z(0)| \sin(\omega_r t + \xi),
\end{equation} \end{subequations}
where $\xi$ is the \textit{complex phase} of $z(0)$.

From the analysis of the quantum formulation of the resonator, it is possible to verify that the probability amplitudes for the flux and charge variables follow Gaussian distributions for the ground state, being normalized to represent a quantum zero-point energy state. These distributions are normalized to ensure that they represent a quantum zero-point energy state accurately. The Gaussian nature of the probability distributions implies that the most probable values of the flux and charge variables cluster around their respective mean values, with decreasing probability as the values deviate further from the mean. This behavior is characteristic of systems in equilibrium.

In practice, the relevance of the quantization procedure for LC oscillators hinges on two critical conditions. First, the oscillator must be effectively isolated from uncontrollable factors to keep its energy levels much narrower than their spacing. This requires a high-quality factor Q for the oscillator. Superconductors like aluminum and niobium, because of their low losses, are suitable for achieving the necessary Q values. The second condition is that the energy gap between adjacent quantum states should exceed thermal energy. Microwave frequency circuits, operated at low temperatures (around 10~mK) in dilution refrigerators, satisfy this condition. With these requirements met, a microwave-range oscillator can operate in the quantum regime. It can be prepared in its ground state $|n = 0\ket$ by allowing it to evolve for a few photon lifetimes~\cite{Blais_2021}.

\subsubsection{Quantum Resonator – Transmission Line}

Understanding the properties of transmission line resonators within the context of cQED is important for optimizing their performance in several applications, such as the exploration of synergistic effects when coupled to other superconducting elements, such as qubits. These interactions can lead to novel phenomena and effects that are not present in isolation. In addition, integrating the discussion of transmission line resonators into the cQED framework facilitates experimental validation of theoretical predictions and opens up possibilities for exploring new regimes of operation in quantum technologies.

In this section, the significance of multimode structures is explored by examining the electromagnetic properties of coplanar waveguide resonators, which are modeled as linear, dispersion-free one-dimensional media. 
\begin{figure}[!ht]
    \centering
    \includegraphics[width = 87mm]{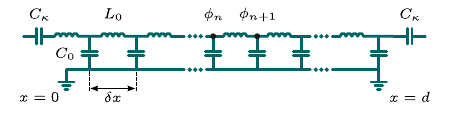}
    \caption{Telegrapher model of an open-ended transmission line resonator of length $d$. $L_0$ and $C_0$ are, respectively, the inductance and capacitance associated to each node $n$ of flux $\phi_n$. The resonator is coupled to external transmission lines (not shown) at its input and output ports via the capacitors $C_\kappa$.}
    \label{fig:quareslines}
\end{figure}

Consider the circuit depicted in Fig. \ref{fig:quareslines}, composed of intercalated \textit{inductances} $L_n$ and \textit{shunt capacitances} $C_n$. The nodes are physically distanced by $\delta x$, adding up to a \textit{total length} of $d$. It is important to visualize that $C_0 = \delta x c_0$ and $L_0 = \delta x l_0$, with the lower case notation for \textit{physical quantity per unit length}. By quantizing fields in transmission lines, it is determined the potential to create resonators with distributed parameters. Thus, the circuit's Hamiltonian using \textit{charge} $Q_n$ and flux $\phi_n$ variables:
\begin{equation}
H = \sum_{n=0}^{N-1} \left[\frac{1}{2C_0}Q_n^2 + \frac{1}{2L_0}(\phi_{n+1} - \phi_n)^2\right].
\end{equation}

Consider the linear charge density as a function of $n\delta x$: 
\begin{equation}
    \lambda(n\delta x) = \frac{Q_n}{\delta x}.
\end{equation} 
Furthermore,  $c_0$ and $l_0$ are the capacitance and inductance per unit length, respectively. In the limit as $\delta x \to 0$, the classical Hamiltonian can be expressed as the following integral:
\begin{equation} \label{eq:quareslinham}
    H = \int_0^d \left(\frac{1}{2c_0}\lambda(x)^2 + \frac{1}{2l_0}(\partial_x\phi(x))^2\right)dx.
\end{equation}
Obtaining the equations of motion derived from Eq. (\ref{eq:quareslinham}) using Hamiltonian classical field theory:
\begin{equation}
    \dot{\lambda} = - \frac{\delta \mathcal{H}}{\delta \phi} = - \frac{1}{l_0}\frac{\partial^2\phi}{\partial x^2},
\end{equation}
\begin{equation}
    \dot{\phi} = \frac{\delta \mathcal{H}}{\delta \lambda} = \frac{1}{c_0}\lambda.
\end{equation}
Considering the wave propagation velocity $v_0$ on the circuit, whose equations lead to the flux wave equation:
\begin{equation}\label{eq:fluxWaveEq}
    v_0^2\frac{\partial^2\phi}{\partial x^2} = \frac{\partial^2\phi}{\partial t^2},
\end{equation}
where:
\begin{equation}
    v_0 = \frac{1}{\sqrt{l_0c_0}}.
\end{equation}
The general solution for $\phi(x, t)$ can be expressed as:
\begin{equation}\label{eq:quaresfluxgen}
    \phi(x, t) = \int_{-\infty}^{\infty} A_k\cos(kx + \alpha_k)\cos(\omega t + \beta_k)dk,
\end{equation}
where $A_k$ is the amplitude, $\alpha_k$ is the spatial phase constant, and $\beta_k$ is the temporal phase constant for each wave component with a specific wave number $k$. Here, the frequency $\omega = v_0k$. From the continuity equation, the open-circuit boundary condition is obtained at the position $x_{oc}$ ($x_{oc}$ represents the physical coordinates of the open ends of the transmission line resonator, such as $x=0$ and $x=d$, where the boundary conditions are applied):
\begin{equation}
    I(x_{oc}, t) = -\frac{1}{l_0}\frac{\partial\phi}{\partial x}\Big|_{x_{oc}} = 0,
\end{equation}
and from the flux definition, the short-circuit boundary condition in $V_{sc}$ is obtained:
\begin{equation}
   V(x_{sc}, t) = \frac{\partial\phi}{\partial t}\Big|_{x_{sc}} = 0.
\end{equation}

In the case of an open-ended half-wavelength transmission line, both ends are in an open circuit, leading to $k = \frac{n\pi}{d}$ and $\alpha_k = 0$. Similarly, for a quarter-wavelength transmission line, each end is either short or open, resulting in $k = \left(n + \frac{1}{2}\right)\frac{\pi}{d}$ and $\alpha_k = 0$. The constants $A_k$ and $\beta_k$ are determined by the initial conditions. Therefore, Eq. (\ref{eq:quaresfluxgen}) can be expressed as: 
\begin{equation}
\phi(x, t) = \sum_n A_n\cos(k_nx)\cos(\omega_nt + \beta_n).
\end{equation}

The separation of variables requires explicit normalization of the spatial eigenfunctions $\cos(k_n x)$ to establish an orthonormal basis. While these eigenfunctions satisfy orthogonality over the resonator length $d$, their normalization is achieved through the introduction of a $\sqrt{2}$ prefactor, yielding the condition:
\begin{equation}
    \int_0^d \left[\sqrt{2}\cos(k_n x)\right] \left[\sqrt{2}\cos(k_m x)\right] dx = \delta_{nm} d.
\end{equation}
This normalization permits the flux field $\phi(x,t)$ to be expressed as a superposition of orthonormal modes. The resulting form:
\begin{equation}
    \phi(x, t) = \sum_n \sqrt{2}A_n \cos(k_n x)\left(\cos(\omega_n t + \beta_n)\right)
\end{equation}
explicitly separates the spatial and temporal dependencies. This decomposition naturally leads to the identification of:
\begin{itemize}
    \item Spatial mode functions: $u_n(x) = \sqrt{2}\cos(k_n x)$
    \item Temporal amplitudes: $\phi_n(t) = A_n\cos(\omega_n t + \beta_n)$
\end{itemize}
establishing the foundation for canonical quantization through the product decomposition $\phi(x,t) = \sum_n u_n(x)\phi_n(t)$. 

The spatial and time-dependent components are defined such that $u_n(x)$ has a length norm $d$ for the transmission line, and $\phi_n(t)$ contains the two constants depending on the initial conditions. It's worth noting that $u_n(t)$ is already fully defined and depends on the propagation mode $n$, while $\phi_n(t)$ remains a functional variable. Inserting $\phi(x, t)$ into the Hamiltonian (Eq. (\ref{eq:quareslinham})) results in a summation of uncoupled harmonic oscillators:
\begin{equation}\label{eq:quaressum}
H = \sum_{n=0}^\infty \left(\frac{1}{2}\frac{Q_n^2}{C_r} + \frac{1}{2}C_r\omega_n^2\phi_n^2\right).
\end{equation} 
Here, $C_r = dC_0$ represents the total capacitance of the line, $Q_n = C_r\dot{\phi}_n(t)$ is the charge conjugate to $\phi_n$, and the angular frequency $\omega_n$ represents the modal frequency of the $n$-th standing wave mode of the transmission line resonator.

Generalizing Eqs. (\ref{eq:phiop}) and (\ref{eq:qop}) for the operators $\hat{\phi}$ and $\hat{Q}$ for each pair $Q_n$ and $\phi_n$ in Eq. (\ref{eq:quaressum}), the quantum Hamiltonian for a transmission line is obtained:  
\begin{equation}
    \hat{H} = \sum_{n = 0}^\infty \hbar \omega_n \left(\hat{a}_n^\dagger\hat{a}_n + \frac{1}{2}\right),
\end{equation} 
which represents a sum of quantum harmonic oscillators, each with frequency $\omega_n$ and ladder operators $\hat{a}_n$ and $\hat{a}^\dagger_n$.

From the analysis of the modeling of the transmission line, it is observed that the behavior of a transmission line can be approximated to a sum of uncoupled harmonic oscillators. This approximation is possible by ignoring the presence of the input and output port capacitors and by considering a homogenous medium forming of the resonator. 

\subsection{Practical Examples of Resonators}
In the realm of cQED, the resonators share fundamental principles related to electromagnetic fields, resonant behavior, and energy storage. Their specific designs and applications may differ, but their common features make them essential tools in various scientific and engineering contexts. Some practical examples can be cited: 

\begin{itemize}
    \item \textbf{Coaxial stub cavities}: A specific type of resonator architecture used in cQED experiments that is designed to address challenges related to losses, cross-talk, and packaging that can limit coherence and scalability in planar devices. Researchers can use this platform to build more intricate circuits with multiple resonators and qubits. In the near term, it allows for significantly more complex, many-resonator, many-qubit circuits~\cite{Axline2016AnAF, Rosenberg2017}.
    
    \item \textbf{Cylindrical Cavities}: a resonant structure with a cylindrical geometry, supporting standing electromagnetic waves at specific frequencies determined by its dimensions and boundary conditions. In cQED, these cavities are vital for creating and manipulating quantum coherence, coupling superconducting qubits to microwave photons, facilitating efficient quantum state transfer, implementing error correction protocols, and simulating quantum systems~\cite{Frunzio2004FabricationAC, paik_observation_2011}.
    
    \item \textbf{CPW resonators}: integral to cQED experiments, utilize coplanar waveguide structures to support standing electromagnetic waves at specific frequencies determined by their dimensions and boundary conditions. Fabricated with superconducting materials like niobium or aluminum, CPW resonators consist of a central conductor, ground planes, and a dielectric layer. Their applications include creating and maintaining quantum coherence, coupling qubits to microwave photons, manipulating quantum states, and implementing error correction protocols~\cite{wallraff_strong_2004}.
\end{itemize}

\section{Transmons}\label{sec:transmon}
The journey toward a fault-tolerant quantum computer began with the first experimental demonstrations of superconducting qubits at the turn of the millennium. These pioneering designs included the charge qubit, or “Cooper-pair box”~\cite{Nakamura1999}, and the flux qubit, which relied on quantum tunneling of charge and magnetic flux, respectively~\cite{Mooij1999}. While foundational, these early qubits suffered from short coherence times due to a strong sensitivity to environmental noise — predominantly charge noise for the charge qubit and flux noise for the flux qubit. The development of subsequent qubit designs was therefore largely driven by the goal of mitigating these decoherence channels.

The transmon is a type of superconducting qubit developed to minimize sensitivity to charge noise. Originating from Yale University in 2007~\cite{Koch_2007}, it's based on the Cooper-pair box design, featuring a Josephson junction parallel to a coplanar shunt capacitor. This shunting capacitor reduces sensitivity to charge noise while maintaining anharmonicity. By increasing the Josephson energy to charging energy ratio, the transmon's energy level spacings become insensitive to offset charge fluctuations.  

Conjointly, transmons are essential components in quantum computing, enabling the creation, manipulation, and storage of quantum coherence~\cite{kjaergaard_superconducting_2020}. Coupling transmons to resonators facilitates efficient quantum state transfer and gate operations, making them valuable tools for advancing quantum information science and computing~\cite{DiCarlo2009}.

Overall, the transmon qubit plays a central role in advancing research and applications in cQED by offering long coherence times, tunable resonance frequencies, strong anharmonicity, compatibility with CPW resonators, and scalability for building practical quantum technologies~\cite{kjaergaard_superconducting_2020}.

\begin{figure}[!ht]
    \centering
    \includegraphics[width = 38mm]{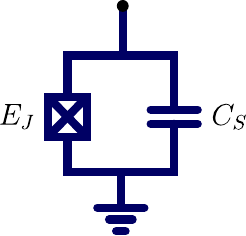}
    \caption{A transmon qubit represented as a circuit element, consisting of a Josephson junction characterized by its energy $E_J$ shunted by a large capacitor $C_S$.}
    \label{fig:transmon}
\end{figure}

In a practical sense, the transmon is built from an LC circuit by replacing an inductor with a Josephson junction. Nonetheless, the design of the transmon is conceived for an operation regime for a significantly increased ratio of Josephson energy and charging energy $E_J/E_C$. Furthermore, its charge dispersion decreases exponentially with this ratio~\cite{Koch_2007}.

For a quantitative description of a transmon, its Hamiltonian can be obtained by combining Eqs. (\ref{eq:resham}) and (\ref{eq:jjham}):
\begin{equation}
    \hat{H} = \frac{\hat{Q}_C^2}{2C} - E_J \cos\left( \frac{\hat{\phi}}{\phi_0}\right) + \frac{\hat{Q}_J^2}{2C_J},
\end{equation}
where $\hat{\phi}$ is the \textit{flux operator}, $\phi_0$ is the \textit{reduced flux quantum} $(\hbar/2e)$, $\hat{Q}_C^2$ is the \textit{operator related to the charge accumulated across the capacitor}, $\hat{Q}_J^2$ is the \textit{operator related to the charge accumulated across the junction} and $C_J$ is the \textit{capacitance}. Defining $\hat{Q} = \hat{Q}_C + \hat{Q}_J$ and $C_{\Sigma} = C + C_J$, the Hamiltonian for this quantum system can be expressed as follows:
\begin{equation}
    \hat{H} = \frac{\hat{Q}^2}{2C_\Sigma} - E_J \cos\left(\frac{\hat{\phi}}{\phi_0}\right),
\end{equation}
defining $\hat{n} = \frac{\hat{Q}}{2e}$ and $\hat{\varphi} = \frac{\hat{\phi}}{\hat{\phi}_0}$, and noting that $\hat{n}$  and $\hbar\hat{\varphi}$ are conjugated variables:
\begin{equation}
    \hat{H} = 4E_C\hat{n}^2 - E_J \cos(\hat{\varphi}),
\end{equation}
where the \textit{charging energy} $E_C$ is given by $E_C = \frac{e^2}{2C\Sigma}$.

The ratio $E_J/E_C$ can control how much the charge and flux values for the qubit's number states disperse. If the charge dispersion is low, the system becomes highly sensitive to charge noise, which can alter the qubit's frequency. To make it more resistant to such noise sources and increase coherence time, it is useful to work in the transmon regime with $E_J/E_C \gg 1$~\cite{Koch_2007}. In this regime, the charge is spread out, that is, the energy eigenstates of the qubit are a superposition of many different charge number states, and the flux dispersion $\bra\hat{\varphi}\ket^2$ is low. The Hamiltonian can then be approximated up to the fourth order of $\hat{\varphi}$ as:
\begin{equation}
    \hat{H} = 4E_C\hat{n}^2 + \frac{E_J}{2}\hat{\varphi}^2 - \frac{E_J}{4!}\hat{\varphi}^4.
\end{equation}

In this condition, the transmon behaves like a weakly anharmonic oscillator. Moreover, we introduce the creation $\hat{b}^\dagger$ and annihilation $\hat{b}$ operators for the elementary energy excitations of the transmon circuit as follows:
\begin{equation}
   \hat{\varphi} = \left(\frac{2E_C}{E_J}\right)^{1/4}(\hat{b}^\dagger + \hat{b}),
\end{equation}
\begin{equation}
    \hat{n} = \frac{i}{2}\left(\frac{E_J}{2E_C}\right)^{1/4}(\hat{b}^\dagger - \hat{b}).
\end{equation}
Expressing the Hamiltonian in terms of $\hat{b}$ and $\hat{b}^\dagger$ and applying the rotating wave approximation (RWA), which is a standard method in quantum optics used to neglect fast-oscillating terms that average to zero over the relevant timescales of the system's dynamics~\cite{Girvin_2011_CQED}, results in the simplified form:
\begin{equation}\hat{H} = \sqrt{8E_C E_J}\hat{b}^\dagger \hat{b} - \frac{E_C}{12}(\hat{b}^\dagger + \hat{b})^4
\label{eq:tranappham} 
\approx \hbar \omega_q \hat{b}^\dagger \hat{b} - \frac{E_C}{2}\hat{b}^\dagger \hat{b}^\dagger \hat{b}\hat{b},
\end{equation}
where $\hbar\omega_q = \sqrt{8E_C E_J} - E_C$. The RWA is valid when $\hbar\omega_q \gg \frac{E_C}{4}$, which is easily satisfied in the transmon regime. The last term in Eq. (\ref{eq:tranappham}) represents a Kerr nonlinearity, with $\frac{E_C}{\hbar}$ acting as the Kerr frequency shift per excitation of the nonlinear oscillator.

Using the commutation relation $\hat{b}\hat{b}^\dagger = \hat{b}^\dagger \hat{b} + \hat{1}$ and neglecting constants, the Hamiltonian simplifies to:
\begin{equation}
    \hat{H} = \hbar\left(\omega_q - \frac{E_C}{2\hbar}\right)\hat{b}^\dagger \hat{b} - \frac{E_C}{2}(\hat{b}^\dagger \hat{b})^2.
\end{equation}

Furthermore, with the excitation operator written in the basis of Fock states, where \( k \) represents the \textit{number of excitations} in the quantum system, specifically in the Fock state basis:
\begin{equation}
    \hat{b}^\dagger\hat{b} = \sum_{k = 0}^\infty = k |k\ket \bra k|.
\end{equation}
The Hamiltonian becomes:
\begin{equation}
    \hat{H} = \sum_k \left[ k \hbar \left(\omega_q - \frac{E_C}{2\hbar} \right)-k^2\frac{E_C}{2}\right]|k\ket\bra k|,
\end{equation}
and, for the \textit{first excitation frequency} $\omega_{01} = \omega_q - E_C/\hbar$, as the transition $|0\ket \to |1\ket$, the Hamiltonian becomes:
\begin{equation}
    \hat{H} = \sum_k \left[ k \hbar \omega_{01}-k(k-1)\frac{E_C}{2}\right]|k\ket\bra k|.
\end{equation}

When considering only the ground and first excited states, the equation can be further simplified to:
\begin{equation}\label{eq:twolevelcoup}
    \hat{H} =\omega_{01}|1\ket\bra 1|= -\frac{w_{01}}{2}\hat{\sigma}_z,
\end{equation}
where $\hat{\sigma}_z$ is the Pauli-Z matrix.

Finally, from the description above of the transmon, some inferences could be made:

\begin{itemize}
    \item The general behavior of the transmon relates to a weakly anharmonic oscillator;
    \item The transmon regime allows the approximation for the operation of the transmon that represents the qubit's number states dispersion;
    \item The second excitation frequency is $w_{12} = w_{01} - E_C$ relates to the lowering caused by the anharmonicity $\alpha = E_C$.
\end{itemize}

\subsection{Transmon-Resonator Coupling}
At this point, the building blocks of an operational qubit are set. To accomplish this, a way to allow the qubit to function effectively is presented: transmon-resonator coupling. This occurs because the coupling above provides nonlinearity, tunable energy levels, quantum coherence, and efficient readout. These properties are essential for performing quantum information processing tasks such as qubit manipulation, storage, measurement, and feedback control in quantum algorithms~\cite{Blais2004}. 

Firstly, some fundamental concepts of cQED will be presented for the analysis of the transmon-resonator coupling. Afterward, these principles will be applied to scenarios where a transmon qubit interacts with either a gate voltage or an external oscillator.

The process of circuit analysis can be outlined as follows:
\begin{enumerate}
    \item Start by applying Kirchhoff's laws, which describe how voltages and currents behave in the circuit;
    \item Reduce the number of variables to the minimum necessary;
    \item Express the capacitive energies in terms of flux derivatives ($E_C = C\dot{\phi}^2/2$) and inductive/Josephson energies in terms of the flux ($E_I = \phi^2/2L$, $E_J = -E \cos(\phi/\phi_0)$);
    \item Formulate the Lagrangian, where capacitive (inductive/Josephson) energies correspond to kinetic (potential) energies;
    \item Finally, derive the Hamiltonian using a Legendre transform.
\end{enumerate}

\subsubsection{Transmon coupled to gate voltage}

\begin{figure}[!ht]
    \centering
    \includegraphics[width = 76mm]{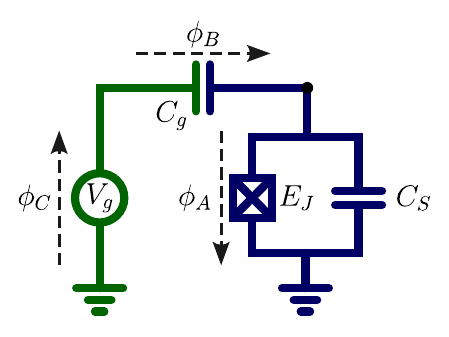}
    \caption{Representation of a transmon coupled to a gate voltage~\cite{Blais_2021}}
    \label{fig:restransgatvol}
\end{figure}

Now, it will be discussed a transmon qubit coupled to a gate voltage, as shown in Fig. \ref{fig:restransgatvol}.

Following the steps of the analysis, according to Kirchhoff's voltage law:
\begin{equation}
    V_g + V_B + V_A = 0 \implies V_g + \dot{\phi}_B + \dot{\phi}_A = 0,
\end{equation}
where $V_g(t)$ is the classical, time-dependent gate voltage applied externally. The terms $V_A$ and $V_B$ represent the internal voltages of the circuit, which are defined by the time-derivatives of the node fluxes $\phi_A$ and $\phi_B$. Rearranging this gives the relation:
\begin{equation}
    \dot{\phi}_B = -\dot{\phi}_A - V_g.
\end{equation}

The energy stored in the capacitor $C_S$ is $C_S \dot{\phi}_A^2/2$, and in the capacitor $C_g$ is $C_g (\dot{\phi}_A + V_g)^2/2$. The inductive energy of the Josephson Junction is $E_J \cos(\phi_A/\Phi_0)$. Thus, the Lagrangian for this system can be expressed as:
\begin{equation}
   \mathcal{L} = \frac{C_S}{2} \dot{\phi}_A^2 + \frac{C_g}{2} (\dot{\phi}_A + \dot{\phi}_C)^2 + E_J \cos(\varphi_A),
\end{equation}
where $E_J = (\phi_0/2\pi)I_c$. The conjugate momentum of $\phi_A$ is $Q_A = \partial_{\dot{\phi}_A} \mathcal{L} = (C_S + C_g) \dot{\phi}_A + C_g V_g$. The Hamiltonian results from the Legendre transform:
\begin{equation}
    H = \dot{\phi}_A Q_A - \mathcal{L} ,
\end{equation}
\begin{equation}
    H = \frac{(Q_A - C_g V_g)^2}{2(C_\Sigma + C_g)} - E_J \cos(\frac{\phi_A}{\phi_0}).
\end{equation}
Expanding the capacitive energy term leads to a drive-transmon interaction term in the form of:
\begin{equation}\label{eq:ham_drive}
    \hat{H}_{\text{drive}} = -\frac{C_g Q_{\text{zpf}}}{2(C_\Sigma + C_g)}(V_g^* - V_g)(\hat{a}^\dagger-\hat{a}).
\end{equation}

At the transmon regime, considering the RWA approximation:
\begin{equation}\label{eq:rwatransmongate}
    \hat{H}_{\text{drive}} = -\frac{C_g Q_{\text{zpf}}}{2(C_\Sigma + C_g)}(V_g\hat{a}^\dagger + V_g^*\hat{a}).
\end{equation}
From the analysis of Eq. (\ref{eq:rwatransmongate}), it is possible to arrive at a scenario where the circuit is represented by an operator involving \( \hat{a} \) and \( \hat{a}^\dagger \).

Furthermore, the RWA is applied to model the time dependence of \( V_g(t) \), where \( V_g(t) = \tilde{V}_{g}e^{-i\omega_d t} \), and \( \omega_d \) closely match the resonant frequency of the transmon. 

When the quantum behavior of the drives is considered negligible, it is possible to treat the drives as classical forces, significantly easing the mathematical treatment of the system, in what is called the stiff pump regime~\cite{reagor2019}.

\subsubsection{Transmon coupled to resonator}
\begin{figure}[!ht]
    \centering
    \includegraphics[width = 76mm]{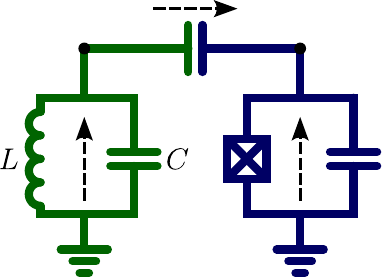}
    \caption{Representation of a transmon coupled to a resonator, with $L$ and $C$ representing the resonator's inductance and capacitance, respectively.~\cite{Blais_2021}.}
    \label{fig:restransres}
\end{figure}

This segment follows the discussion on the transmon coupled to a resonator. The depiction of the aforementioned can be seen in Fig. \ref{fig:restransres}.

Following a method analogous to the previously discussed case of the gate-coupled transmon, the analysis for the capacitive interconnection between a transmon and a resonator begins with Kirchhoff's voltage law. Designating $V_1$, $V_2$, and $V_3$ as the voltages across the resonator, gate, and transmon, correspondingly, with orientations structured such that the Kirchhoff voltage law is expressed as:
\begin{equation}
    V_1 + V_2 + V_3 = 0.
\end{equation}

Embracing the flux notation:
\begin{equation}
    \dot{\phi}_1 + \dot{\phi}_2 + \dot{\phi}_3 = 0 \implies \dot{\phi}_2 = -\dot{\phi}_1 - \dot{\phi}_3.
\end{equation}
Reducing the number of variables to the minimum necessary and expressing the capacitive energies in terms of flux derivatives and inductive/Josephson energies, the Lagrangian of this system is delineated as:
\begin{eqnarray}
\mathcal{L} = \frac{C_T}{2}\dot{\phi}_3^2 + \frac{C_R}{2}\dot{\phi}_1^2 + \frac{C_g}{2}(\dot{\phi}_1 + \dot{\phi}_3)^2   +\nonumber\\ 
+ E_J \cos \left(\frac{\phi_3}{\phi_0}\right) - \frac{\phi_1^2}{2L_R},
\end{eqnarray}
here, $C_g$ stands for the coupling capacitance, $C_R$ and $L_R$ denote the resonator's capacitance and inductance, respectively, $C_T$ and $E_J$ represent the capacitance and Josephson junction's characteristic energy of the transmon, and $\phi_0$ signifies the flux quantum.

The conjugate momenta corresponding to $\phi_1$ and $\phi_3$ are derived as follows:
\begin{equation}
\left\{ 
    \begin{array}{cc}
    Q_1 &= \dfrac{\partial L}{\partial \dot{\phi}_1} = (C_R + C_g)\dot{\phi}_1 + C_g\dot{\phi}_3  \\ & \\
    Q_3 &= \dfrac{\partial L}{\partial \dot{\phi}_3} = (C_T + C_g)\dot{\phi}_3 + C_g\dot{\phi}_1 
    \end{array}
\right. .
\end{equation}
Consequently, expressing $\dot{\phi}_1$ and $\dot{\phi}_3$ in terms of $Q_1$ and $Q_3$ yields:
\begin{equation}
\left\{
    \begin{array}{cc}
    \dot{\phi}_1 &= \dfrac{C_T + C_g}{D}Q_1 - \dfrac{C_g}{D}Q_3  \\ & \\
    \dot{\phi}_3 &= \dfrac{C_R + C_g}{D}Q_3 - \dfrac{C_g}{D}Q_1 
    \end{array}
\right.,
\end{equation}
where $D = C_R C_T + C_R C_g + C_T C_g$. From the Legendre transform, the Hamiltonian of the system can then be expressed in terms of $\phi_1$, $\phi_3$, $Q_1$, and $Q_3$:
\begin{equation}
\hat{H} = \dot{\phi}_1 Q_1 + \dot{\phi}_3 Q_3 - \mathcal{L} = \hat{H}_T + \hat{H}_R + \hat{H}_g.
\end{equation}
Here, $\hat{H}_T$, $\hat{H}_R$, and $\hat{H}_g$ signify the transmon, resonator, and coupling terms, respectively, defined as (noting that $Q = 2en$):
\begin{align}
\hat{H}_T &= (4e^2)\frac{C_R + C_g}{2D}\hat{n}_T^2 - E_J \cos \varphi_T, \\
\hat{H}_R &= (4e^2)\frac{C_T + C_g}{2D}\hat{n}_R^2 + \frac{\phi_0^2}{2L_R}\varphi_R^2,  \\
\hat{H}_g &= -(4e^2)\frac{C_g}{D}\hat{n}_T\hat{n}_R. 
\end{align}

These equations represent the Hamiltonian terms for different components in the system, where a transmon is coupled to a resonator. Analyzing these equations, some inferences can be made:

\begin{itemize}
    \item \( \hat{H}_T \) is the Hamiltonian term associated with the transmon, which relates to the charging energy of the transmon due to its capacitance with the resonator and gate and the Josephson energy of the transmon;
    \item  \( H_R \) represents the Hamiltonian term associated with the resonator, which relates to the charging energy of the resonator due to its capacitance with the transmon and gate and the kinetic energy of the resonator;
    \item \( H_g \) signifies the Hamiltonian term associated with the coupling between the transmon and the resonator, which represents their interaction due to their capacitive coupling;
    \item $\hat{n}_T$ and $\hat{n}_R$ represent charge number operators for the transmon and for the resonator, respectively.
\end{itemize}

Moreover, the negative sign in $H_g$ aligns with conventional notation. Expressing the Hamiltonian in terms of creation and annihilation operators $\hat{a}$, $\hat{a}^{\dagger}$, $\hat{b}$, and $\hat{b}^{\dagger}$ for the resonator and transmon, respectively, and approximating the cosine function up to the fourth power of the reduced flux, we arrive at:
\begin{equation}\hat{H} = \hbar\omega_R \hat{a}^{\dagger}\hat{a} + \hbar\omega_T \hat{b}^{\dagger}\hat{b} - \frac{E_C}{2}\hat{b}^{\dagger}\hat{b}^{\dagger}\hat{b}\hat{b} \label{eq:hamtranscoupledres}
- \hbar g(\hat{b}^\dagger - \hat{b})(\hat{a}^\dagger - \hat{a}),
\end{equation}
where $\omega_R$ is the resonant angular frequency of the linear LC resonator and $\omega_T$ is the transition angular frequency of the transmon qubit.
The transmon-resonator coupling constant is determined by the expression:
\begin{equation}
g = \omega_r \frac{C_g}{C_T} \left( \frac{E_J}{2E_C} \right)^{\frac{1}{4}} \sqrt{\frac{\pi Z_R}{2 R_K}},
\end{equation}
where the terms \(Z_R = \sqrt{L_R/C_R}\) and \(R_K = h/e^2\) represent the characteristic impedance and quantum resistance, respectively.

It is important to note that the derivation presented here pertains to a scenario involving a single resonator electromagnetic mode. Multiple modes are common in practical setups where the transmon is coupled to a microwave cavity or a transmission line. However, one of these modes is often specifically engineered to exhibit strong coupling with the qubit. Consequently, the weakly interacting modes can be disregarded, simplifying the analysis significantly. However, the study of weakly interacting modes within quantum system environments, such as Feshbach-Fano partitioning~\cite{Schuray2017}, is a field of research with ongoing developments~\cite{Kockum_2018}.

\subsubsection{Rotating frames and approximations}
The derivation of the coupled Hamiltonian, Eq. (\ref{eq:hamtranscoupledres}), presented remains exact. In regimes typical of transmon-cavity coupling scenarios, certain approximations become applicable, enabling analytical solutions. Commencing with the assumption of the two-level system approximation for the transmon, as described by Eq. (\ref{eq:twolevelcoup}), the Hamiltonian assumes the form:
\begin{equation}\label{eq:gen_ham_code}
\hat{H} = \hbar\omega_R \hat{a}^\dagger \hat{a} - \frac{\hbar\omega_T}{2}\hat{\sigma}_z - \underbrace{\hbar g(\hat{\sigma}_+ - \hat{\sigma}_-)(\hat{a}^\dagger - \hat{a})}_{\hat{H}_g},
\end{equation}
where $\hat{H}_g$ is the coupling term of the Hamiltonian, and $\hat{\sigma}_+$ and $\hat{\sigma}_-$ are the Pauli raising and lowering operators. Within the two-level approximation, where only the transmon's ground $|0\rangle$ and first excited $|1\rangle$ states are considered, these operators are equivalent to the transmon's creation and annihilation operators, $\hat{b}^\dagger$ and $\hat{b}$, respectively:
\begin{equation}
\left\{
    \begin{array}{ccc}
    \hat{\sigma}_+ \equiv& |1\rangle\langle 0| \approx& \hat{b}^\dagger \\
    \hat{\sigma}_- \equiv& |0\rangle\langle 1| \approx& \hat{b}
    \end{array}
\right..
\end{equation}

The initial two terms pertain to the excitation numbers of the resonator and transmon and can be disregarded through the application of a unitary transformation:
\begin{equation}
\hat{U}(t) = \exp{\left( -i\omega_R t \hat{a}^\dagger \hat{a} + i\frac{\omega_T t}{2}\hat{\sigma}_z \right)},
\end{equation}
this transformation relocates the problem to an appropriate interaction picture. The two operators summed in the exponent mutually commute, allowing the interpretation of \( U(t) \) as a composition of two independent unitary transformations:
\begin{equation}
\hat{U}(t) = \hat{U}_R (t) \hat{U}_T (t) = \hat{U}_T (t) \hat{U}_R (t),
\end{equation}
with
\begin{equation}
\left\{
\begin{array}{cc}
\hat{U}_T(t) &= \exp{\left(i\dfrac{\omega_T t}{2}\hat{\sigma}_z\right)}, \\ & \\
\hat{U}_R(t) &= \exp{\left(-i\omega_R t \hat{a}^\dagger \hat{a}\right)}.
\end{array}
\right. 
\end{equation}

In this newly defined reference frame, the coupling term $\hat{H}_g$ in Eq. (\ref{eq:hamtranscoupledres}) undergoes transformation as:
\begin{equation}\label{eq:rotintconstr}
\hat{H}_{I,g} = \hat{U}^\dagger(t)\hat{H}_g \hat{U}(t).
\end{equation}
This defines the dynamics of the system via:
\begin{equation}\label{eq:schrointrot}
i\hbar\frac{\partial}{\partial t}|\psi_i\rangle = \hat{H}_{I,g}(t)|\psi_i\rangle.
\end{equation}

To comprehend the impact of the coupling Hamiltonian on the evolution of the system's states, it is possible to examine each of the terms of the expansion of the coupling term:
\begin{equation} \label{eq:hgexpansion}
\hat{H}_g = -\hbar g(\hat{a}\hat{\sigma}_+ + \hat{a}^\dagger\hat{\sigma}_- -\hat{a}^\dagger\hat{\sigma}_+ - \hat{a}\hat{\sigma}_-).
\end{equation}

Each component of the interaction Hamiltonian consists of a product of ladder operators associated with the resonator and the transmon, respectively. These operators act on distinct Hilbert spaces and therefore commute, a property that permits the factorization of their action within the full system's tensor product space. As a consequence, any operator confined to the resonator subspace remains invariant under unitary transformations acting solely on the transmon subspace, and vice versa. 

This commutation property not only facilitates analytical tractability but also clarifies the nature of the coupling: each term in the Hamiltonian corresponds to a specific excitation exchange process between the two systems. For instance, the presence of a term involving the annihilation operator of the resonator and the creation operator of the transmon explicitly describes the transfer of a single quantum of excitation from the resonator to the transmon. Conversely, a term featuring the resonator's creation operator and the transmon's annihilation operator signifies the reverse process. This structure makes evident the quantized exchange of excitations that underpins the coherent dynamics between the resonator and the transmon.

Thus, computing the transformation detailed in Eq. (\ref{eq:schrointrot}) to elucidate the impact on each ladder operator. Specifically, the transformation of the lowering operator of the cavity is expressed as:
\begin{equation}
\hat{a}_I(t) = \hat{U}^\dagger \hat{a}\hat{U} = \hat{U}^\dagger_R \hat{a}\hat{U}_R \hat{U}^\dagger_T\hat{U}_T,
\end{equation}
which simplifies to, after expanding in Taylor series:
\begin{equation}
\hat{a}_I(t) = e^{i\omega_R t \hat{a}^\dagger \hat{a}}\hat{a}e^{-i\omega_R t \hat{a}^\dagger \hat{a}} 
=e^{i\omega_R t \hat{a}^\dagger \hat{a}}\hat{a}\sum_{n=0}^{\infty}\frac{(-i\omega_R t)^n}{n!} (\hat{a}^\dagger \hat{a})^n.
\end{equation}
Subsequently, using the commutation relation $[\hat{a}, \hat{a}^\dagger]=1$:
\begin{flalign}
\hat{a}(\hat{a}^\dagger\hat{a})^n = (1 + \hat{a}^\dagger\hat{a})\hat{a}(\hat{a}^\dagger\hat{a})^{n-1} =
\dots = (1 + \hat{a}^\dagger\hat{a})^n \hat{a}.
\end{flalign}
Returning the summation to its exponential form, the transformation for the lowering operator of the cavity \(\hat{a}_I(t)\) is given by:
\begin{equation}
\hat{a}_I(t) = e^{i\omega_R t \hat{a}^\dagger \hat{a}} e^{-i\omega_R t(1 + \hat{a}^\dagger \hat{a})}\hat{a} = e^{-i\omega_R t}\hat{a}.
\end{equation}
As anticipated, the transformation for the raising operator is analogously obtained:
\begin{equation}
\hat{a}^\dagger_I(t) = e^{i\omega_R t}\hat{a}^\dagger.
\end{equation}

The calculation for \( \hat{\sigma}_+ \) follows a similar procedure, incorporating the commutation rules \([\hat{\sigma}_\pm, \hat{\sigma}_z] = \mp\hat{\sigma}_\pm\), resulting in the expression:
\begin{equation}
\hat{\sigma}_z \hat{\sigma}_\pm = \hat{\sigma}_\pm (\hat{\sigma}_z \pm 1)^n.
\end{equation}

For the interaction representation, the transformed operators \( \hat{\sigma}_{I,\pm} I(t) \) are expressed as:
\begin{equation}
\hat{\sigma}_{I,\pm} (t) = e^{\pm i\omega_T t}\hat{\sigma}^\pm.
\end{equation}

Continuing from the transformations of ladder operators, constructing the transformation of the coupling Hamiltonian from Eq. (\ref{eq:rotintconstr}):
\begin{equation}-\frac{1}{\hbar g}\hat{H}_{I,g} = \hat{a}_I\hat{\sigma}_{I,+} + \hat{a}_I^\dagger\hat{\sigma}_{I,-} 
- \hat{a}_I^\dagger\hat{\sigma}_{I,+} - \hat{a}_I\hat{\sigma}_{I,-},
\end{equation}
\begin{equation}-\frac{1}{\hbar g}\hat{H}_{I,g} = \hat{a}\hat{\sigma}_+ e^{-i\Delta t} + \hat{a}^\dagger\hat{\sigma}_- e^{i\Delta t}  
\label{eq:explainRWA}
- \hat{a}^\dagger\hat{\sigma}_+ e^{i\Sigma t} - \hat{a}\hat{\sigma}_- e^{-i\Sigma t},
\end{equation}
here, \( \Delta = \omega_R - \omega_T \) and \( \Sigma = \omega_R + \omega_T \). Eq. (\ref{eq:explainRWA}) serves as the initial point for the widely used but infrequently explained Rotating Wave Approximation (RWA). This approximation relies on the observation that when \( \omega_R \) and \( \omega_T \) are reasonably close, the last two terms rotate at a high frequency and can thus be neglected from the Hamiltonian.

Returning to the dynamics of the interaction picture (Eq. (\ref{eq:schrointrot})), expressing the time-dependence of the state as the action of a propagator (\( |\psi(t)\ket = U(t)|\psi_0\ket \)), the resulting equation to solve is:
\begin{equation}
i\hbar\frac{\partial}{\partial t}\hat{U}(t) = \hat{H}_{I,g}(t)\hat{U}(t).
\end{equation}

Since the interaction Hamiltonian is time-dependent, the propagator does not exhibit a straightforward exponential behavior. Instead, its solution can be attained using the Dyson series:
\begin{eqnarray}
\hat{U}(t) &=& 1 - i \int_{0}^{t} dt_1 \hat{H}_{I,g}(t_1) +\nonumber\\ \label{eq:dysonseries}
&+&(-i^2) \int_{0}^{t} dt_1 \int_{0}^{t_1} dt_2 \hat{H}_{I,g}(t_1)\hat{H}_{I,g}(t_2) + \dots
\end{eqnarray}

In Eq. (\ref{eq:dysonseries}), the rationale behind the RWA becomes apparent. When a rapidly rotating term is included in the Dyson series, it undergoes integration one or more times before contributing to the state's evolution. Integration acts as a low-pass filter, damping high frequencies. Formally:
\begin{equation}
\int dt e^{i\omega t} = \frac{1}{i\omega}e^{i\omega t} \xrightarrow{\omega \to \infty} 0.
\end{equation}
Hence, the last two terms of Eq. (\ref{eq:hgexpansion}) can be safely neglected. In fact, the RWA can be applied before transitioning to the interaction picture, yielding the Hamiltonian:
\begin{equation}\label{eq:jaynes_pre_ham}
\hat{H} = \hbar \omega_R \hat{a}^\dagger \hat{a} - \frac{\hbar\omega_T}{2}\hat{\sigma}_z + \hbar g(\hat{a} \hat{\sigma}_+ + \hat{a}^\dagger \hat{\sigma}_-).
\end{equation}
The exact diagonalization of this Hamiltonian results in a set of states and eigenenergies denominated Jaynes-Cummings spectrum.

\subsubsection{Jaynes-Cummings Spectrum}
The Hamiltonian represented by Eq. (\ref{eq:jaynes_pre_ham}) elucidates the interplay between the resonator and transmon systems. Each term within the Hamiltonian holds significance in terms of the stored excitations within the respective systems. The initial two terms distinctly correspond to the resonator and cavity number operators. In contrast, the coupling term delineates a number-conserving exchange of excitations. The constancy of the overall excitation count dictates that transitions or couplings solely transpire among states adhering to this fundamental rule.

The composite state $|n, s\ket$, where $n$ denotes the number of photons in the resonator mode and $s \in \{g, e\}$ labels the transmon in its ground or excited state, forms a natural basis for describing the coupled dynamics of the system. Under the rotating wave approximation and within the dispersive regime, the total Hamiltonian can be effectively decomposed into an infinite set of approximately decoupled two-level subspaces. 

Each subspace, labeled by an integer $n$, captures the dynamics of a fixed total excitation number $n+1$ and is governed by an effective Hamiltonian $\hat{H}_n$ defined in the subspace spanned by $\{ |n+1, g\ket, |n, e\ket \}$. This reduction relies on the assumption that off-resonant couplings to states such as $|n+2, g\ket$ and $|n-1, e\ket$ are sufficiently suppressed and can be neglected to leading order. Although, in principle, the interaction Hamiltonian allows transitions between any pair of states differing by a single excitation, energy conservation and the weak coupling regime justify the truncation to nearest-neighbor excitation manifolds. Matrix notation represents as:
\begin{equation}
\hat{H}_n = \begin{pmatrix}
\langle n + 1, g | \hat{H} | n + 1, g \rangle & \langle n + 1, g | \hat{H} | n, e \rangle \\ & \\
\langle n, e | \hat{H} | n + 1, g \rangle & \langle n, e | \hat{H} | n, e \rangle
\end{pmatrix}, 
\end{equation}
which results in an effective Hamiltonian that takes the form:
\begin{equation}
\hat{H}_n = \begin{pmatrix}
(n + 1)\omega_R - \dfrac{1}{2}\omega_T  & -g\sqrt{n + 1} \\
-g\sqrt{n + 1} & n\omega_R + \dfrac{1}{2}\omega_T
\end{pmatrix},
\end{equation}
where $\omega_R$ and $\omega_T$ are the bare frequencies of the resonator and transmon, respectively, and $g$ characterizes the coupling strength. This matrix formulation highlights the hybridization between the states $|n+1, g\ket$ and $|n, e\ket$ due to the interaction, leading to dressed eigenstates whose energy splitting depends nonlinearly on the excitation number $n$.

The associated eigenstates and eigenenergies can be readily derived. Hence, defining the detuning $\delta = \omega_R - \omega_T$, $\Omega_n = 2g\sqrt{n + 1}$, and \(\Delta_n = \sqrt{\delta^2 + \Omega_n^2}\). Additionally, introducing the trigonometric functions: 
\begin{subequations}
\begin{align}
    \sin(\Theta_n) &= \dfrac{\Omega_n}{\sqrt{({\Delta_n - \delta})^2 + \Omega_n^2}},
\end{align}
\begin{align}
    \cos(\Theta_n) &= \dfrac{\Delta_n - \delta}{\sqrt{(\Delta_n - \delta)^2+\Omega_n^2}}.
\end{align}
\end{subequations}
The higher-energy state $|n, +\ket$ is defined as:
\begin{equation}
    |n, +\rangle = \cos(\Theta_n) |n + 1, g\rangle - \sin(\Theta_n) |n, e\rangle,
\end{equation}
with an associated energy value of $E_{n,+}$:
\begin{equation}
   E_{n,+} =\left(n + \frac{1}{2}\right)\omega_R + \frac{1}{2}\Delta_n.
\end{equation}

Similarly, the lower-energy state $|n, -\ket$ is defined as:
\begin{equation}
    |n, -\rangle = \sin(\Theta_n) |n + 1, g\rangle + \cos(\Theta_n) |n, e\rangle,
\end{equation}
with an associated energy value of $E_{n,-}$:
\begin{equation}
  E_{n,-}  = \left(n + \frac{1}{2}\right)\omega_R - \frac{1}{2}\Delta_n.
\end{equation}

This analytical derivation of the Jaynes-Cummings spectrum illuminates the hybridization of states $|n + 1, g\ket$ and $|n, e\ket$ within the coupled system. This hybridization leads to mixed states $|n, \pm\ket$ with an energy split characterized by $\Delta_n$. The contributions of either the transmon or resonator in these hybridized states are quantified by $\sin(\Theta_n)$ and $\cos(\Theta_n)$.

Further examination in various regimes sheds light on the implications of this hybridization. In scenarios where both systems are resonant ($\delta = 0$), $\Omega_n = \Delta_n$, resulting in an even distribution of excitation between the transmon and the cavity. Conversely, in the absence of coupling ($g = 0$), the states directly correlate to the subsystem with higher frequency.

Lastly, when a considerable detuning $\delta$ is present in the coupled system, akin to the limit $g \to 0$, the states exhibit tendencies towards dehybridization. This transition minimizes the exchange of excitations between the transmon and resonator systems. Nevertheless, the population of excitation within each subsystem induces alterations in the other's frequency, recognized as the dispersive shift. This condition of substantial $\delta$ is commonly referred to as the dispersive regime, prompting an investigation into the derivation of the effective Hamiltonian for this specific case.

\subsection{Drives and driven systems}
The preceding analysis has focused on resonators, transmons, and the nature of their mutual coupling. However, a comprehensive description of quantum hardware also requires the inclusion of external control and measurement mechanisms, particularly those associated with driven interactions. A driven interaction refers to the coherent modulation of a quantum system's dynamics through an externally applied time-dependent field, typically implemented via a transmission line coupled to the system. Such interactions enable both the initialization and manipulation of quantum states, as well as the extraction of information through measurement protocols.

This section introduces the formal treatment of a transmission line, modeled as a continuum of modes that can be effectively approximated as an extended one-dimensional resonator. The transmission line serves as the conduit for control pulses and measurement signals, mediating the exchange of energy and information between the quantum system and the classical control infrastructure. Through this framework, the system's dynamics can be steered by tailored drive fields, thereby enabling precise quantum operations and state readout essential for quantum information processing.

Considering the Hamiltonian for a quantum transmission line presented in Eq.~\ref{eq:quareslinham} and the resulting flux wave equation in Eq.~\ref{eq:fluxWaveEq}, a form of representing the general solution in terms of propagating waves along the x-axis is:
\begin{equation}\label{eq:gen_solution_flux}
    \phi(x, t) = \phi_L\left(t + \frac{x}{v}\right) + \phi_R\left(t - \frac{x}{v}\right).
\end{equation}
Here, $\phi_L$ and $\phi_R$ correspond to left and right-propagating waves along the x-axis.

To write the expressions compactly, let the index $j \in \{L, R\}$ denote the left- and right-propagating modes, and let $s_j$ be a sign factor where $s_L = +1$ and $s_R = -1$. Restricting Eq. (\ref{eq:gen_solution_flux}) to $0 < x < L$, where $L$ is a \textit{length} arbitrarily large, and assuming periodic boundary conditions, leads to the general solution:
\begin{equation}
    \phi_j(t + s_j x/v) = \sum_{n = 0}^\infty \left[A_{j, n}e^{-i\frac{2n\pi v}{L}(t + s_j x/v)} + A_{j, n}^*e^{i\frac{2n\pi v}{L}(t + s_j x/v)}\right].
\end{equation}
Ensuring real values for the flux involves algebraic simplifications and considering the limit $L \to \infty$, allowing the replacement of the summation with integration, where $\omega_n = 2n\pi v/L$:
\begin{eqnarray}\label{eq:general_solution_fl}
    \phi_j(t + s_j x/v) &=& \int_{0}^{\infty} d\omega \left[A_j(\omega)e^{-i\omega(t + s_j x/v)} + \right. \nonumber \\
    && \left. + A_j^*(\omega)e^{i\omega(t + s_j x/v)}\right]. 
\end{eqnarray}
The charge density, as the canonical conjugate pair of the flux, can be obtained via the relation $\lambda_j = c_0\partial_t\phi_j$. Explicitly:
\begin{eqnarray}
    \lambda_j(t + s_j x/v) &=& -ic_0 \int_{0}^{\infty} d\omega \, \omega \left[A_j(\omega)e^{-i\omega(t + s_j x/v)} - \right. \nonumber \\
    && \left. - A_j^*(\omega)e^{i\omega(t + s_j x/v)}\right].
    \label{eq:charge_density}
\end{eqnarray}

Utilizing Eqs. (\ref{eq:general_solution_fl}) and (\ref{eq:charge_density}), the Hamiltonian (Eq.~\ref{eq:quareslinham}) can be expressed in terms of the coefficients $A_j(\omega)$ and $A_j^*(\omega)$. The outcome, after promoting the coefficients to operators and absorbing normalization factors, is:
\begin{equation}\label{eq:hamiltonian_final}
    \hat{H} = \sum_{j \in \{L,R\}} \int_{0}^{\infty} d\omega \, \hbar \omega \, \hat{A}_j^\dagger(\omega) \hat{A}_j(\omega).
\end{equation}

Defining bosonic creation operators $\hat{b}_j^\dagger(\omega)$ that are proportional to $\hat{A}_j^\dagger(\omega)$, the Hamiltonian takes the standard form for a continuum of harmonic oscillators:
\begin{equation}
    \hat{H} = \sum_{j \in \{L,R\}} \int_{0}^{\infty} d\omega \, \hbar \omega \, \hat{b}^\dagger_{j}(\omega) \hat{b}_{j}(\omega),
    \label{eq:hamiltonian_bosonic}
\end{equation} 
where the bosonic commutation relations for the operators are assumed: $[\hat{b}_j (\omega) , \hat{b}_{j'}^\dagger (\omega')] = \delta_{j,j'} \delta(\omega-\omega')$. The integral can be discretized, allowing the interpretation of the transmission line as a collection of harmonic oscillators with frequencies $\omega_n$:
\begin{equation}\label{eq:hamiltonian_harmonic_osc}
    \hat{H} = \sum_{n=0}^{\infty} \sum_{j \in \{L,R\}} \hbar \omega_n \hat{b}^\dagger_{j,n} \hat{b}_{j,n}.
\end{equation}

It's noteworthy that for discrete frequencies, $[\hat{b}_{j,n}, \hat{b}^\dagger_{j',n'}] = \delta_{j,j'}\delta_{n,n'}$. Constant offsets to the Hamiltonian were disregarded in both expressions.

Having defined the coefficients $A_j(\omega)$ in terms of bosonic operators, it is possible to reformulate the flux wave expression~\cite{Blais_2021}:
\begin{eqnarray}
    \phi_j(t + s_j x/v) &=& \int_{0}^{\infty} d\omega \sqrt{\frac{\hbar}{4\pi \omega c_0 v}}\left(\hat{b}_{j}(\omega) e^{-i\omega(t + s_j x/v)} +\right. \nonumber \\
    && \left. +\hat{b}_{j}^\dagger (\omega) e^{i\omega(t + s_j x/v)}\right).
    \label{eq:flux_wave}
\end{eqnarray}
From which the voltage expression can also be derived as $V(x,t) = \partial_t\phi(x,t)$:
\begin{eqnarray}
    V_j(t + s_j x/v) &=& -i\int_{0}^{\infty} d\omega \sqrt{\frac{\hbar \omega}{4\pi c_0 v}}\left(\hat{b}_{j}(\omega) e^{-i\omega(t + s_j x/v)} - \right. \nonumber \\
    && \left. - \hat{b}_{j}^\dagger (\omega) e^{i\omega(t + s_j x/v)}\right).
    \label{eq:voltage_expression}
\end{eqnarray}

The formulation presented above establishes a quantum field-theoretic model of a transmission line and its coupling to superconducting quantum systems. By recasting the electromagnetic field in terms of propagating bosonic modes, a foundational framework is developed for understanding how external signals induce coherent transitions in qubit-resonator systems. This treatment underpins the theoretical basis for qubit control via microwave drives and paves the way for a quantitative analysis of input-output theory, dissipation, and quantum-limited amplification. The transmission line, therefore, emerges not merely as a conduit for electromagnetic energy, but as a central component in the architecture of scalable quantum information platforms.

\subsubsection{Drive connected to a quantum system}

In the scenario where a transmission line emulates a bath of harmonic oscillators, the coupling of this line to a system of interest can be addressed within the framework of open quantum systems. The analysis of signals transmitted and received through such lines involves the application of input-output theory and the utilization of the Langevin equation, which is a stochastic differential equation that describes how a system evolves when subjected to a combination of deterministic and fluctuating (random) forces~\cite{Gliklikh1997The}.

The standard model for such a setup considers a quantum system of interest coupled to an environment, or “bath”, which is represented as a collection of harmonic oscillators. The complete Hamiltonian describing this entire setup is given by:
\begin{equation}  
    \hat{H} = \hat{H}_s + \hat{H}_b + \hat{H}_{int},
\end{equation}
where each term has a distinct physical meaning. Fully expanded, the Hamiltonian is:
\begin{equation}
    \hat{H} = \hat{H}_s + \sum_{n} \hbar\omega_n \hat{b}_n^\dagger \hat{b}_n + \sum_{n} \hbar\left(g_n \hat{a}^\dagger \hat{b}_n + g_n^* \hat{b}_n^\dagger \hat{a}\right).
\end{equation}

The three components of the Hamiltonian are interpreted as follows:
\begin{itemize}
    \item \textbf{$\hat{H}_s$} is the Hamiltonian for the \textbf{system of interest} (e.g., a qubit or transmon) in isolation. It governs the internal energy levels and dynamics of the quantum system itself.

    \item \textbf{$\hat{H}_b = \sum_{n} \hbar\omega_n \hat{b}_n^\dagger \hat{b}_n$} is the Hamiltonian of the \textbf{bath}. It models the environment as a collection of independent harmonic oscillators, where $\hat{b}_n^\dagger$ and $\hat{b}_n$ are the creation and annihilation operators for an excitation in the $n$-th oscillator mode with frequency $\omega_n$.

    \item \textbf{$\hat{H}_{int} = \sum_{n} \hbar\left(g_n \hat{a}^\dagger \hat{b}_n + g_n^* \hat{b}_n^\dagger \hat{a}\right)$} is the \textbf{interaction Hamiltonian}. This term describes the coupling between the system and the bath, allowing for the exchange of energy. It includes terms where the system loses an excitation (via its annihilation operator $\hat{a}$) while a bath mode gains one (via $\hat{b}_n^\dagger$), and the corresponding conjugate process. The parameter $g_n$ is the coupling strength between the system and the $n$-th mode of the bath.
\end{itemize}

The system's dynamics, approached in the Heisenberg picture, entail the assumption of time-dependent operators $\hat{a}(t)$, $\hat{b}_n(t)$, and the resolution of the respective equations:
\begin{eqnarray}
        \dfrac{d}{dt}\hat{a} =& \dfrac{i}{\hbar}[\hat{H}, \hat{a}] , \\ 
        \dfrac{d}{dt}\hat{b}_n =& \dfrac{i}{\hbar}[\hat{H}, \hat{b}_n].
\end{eqnarray}

For $\hat{b}_n$, this translates into:
\begin{equation}
    \frac{d}{dt}\hat{b}_n = -i\omega_n \hat{b}_n - ig_n^* \hat{a}.
\end{equation}

Given the initial condition $\hat{b}_n(t = 0) = \hat{b}_{n}^0$, the solution to the equation of motion is:
\begin{equation}\label{eq:drives_bn_def}    \hat{b}_n(t) = \hat{b}_{n}^0 e^{-i\omega_n t} 
-ig_n^* \int_{0}^{t} dt' \hat{a}(t') e^{-i\omega_n(t - t')} .
\end{equation}
The initial term describes the unimpeded evolution of the bath modes, while the subsequent term arises due to interaction with the system. Substituting this solution back into the equation of motion for $\hat{a}$:
\begin{equation} \label{eq:motion_a_bath} 
\frac{d}{dt}\hat{a} = \frac{i}{\hbar}[\hat{H}_s, \hat{a}] - \sum_{n} i g_n \hat{b}_{n}^0 e^{-i\omega_n t}    
- \sum_{n} |g_n|^2 \int_{0}^{t} dt' \hat{a}(t') e^{-i\omega_n(t - t')}.
\end{equation}

Several approximations prove useful in simplifying Eq. (\ref{eq:motion_a_bath}). The latter term can be reformulated as:
\begin{flalign}\label{eq:drives_approx_reform}
\sum_{n} |g_n|^2 \int_{0}^{t} dt' \hat{a}(t') e^{-i\omega_n(t - t')}   
 = \int_{0}^{t} dt' \left [\hat{a}(t') e^{-i\omega_c(t - t')}\right] \times \nonumber\\
\times\left[ \sum_{n} |g_n|^2 e^{-i(\omega_n - \omega_c)(t - t')}\right],
\end{flalign}
where $\omega_c$ serves as a central frequency resonating with the free evolution of $\hat{a}(t)$, aiming to render the first bracket's term a slowly rotating one. The term in the second bracket denotes the kernel:
\begin{equation}
    K(t - t_0) = \sum_{n} |g_n|^2 e^{-i(\omega_n - \omega_c)(t - t_0)}.
\end{equation}
This, along with a resolution of unity involving Dirac deltas, equates to:
\begin{eqnarray} \label{eq:drives_kernel_diracs}
K(t - t_0) = \int_{-\infty}^{\infty} \frac{d\omega}{2\pi} \left[2\pi \sum_{n} |g_n|^2 \delta(\omega_c + \omega - \omega_n) \right] \times\nonumber\\ \times e^{-i(\omega_n - \omega_c)(t - t_0)}.
\end{eqnarray}

The term within brackets in Eq. (\ref{eq:drives_kernel_diracs}) can be interpreted as the spectral density of the coupling, which defines a frequency-dependent rate. Let's define this rate as a function of a general frequency $\omega'$:
\begin{equation}\label{eq:kappa_def}
\kappa(\omega') \equiv 2\pi \sum_{n} |g_n|^2 \delta(\omega' - \omega_n).
\end{equation}
This function describes how strongly the system couples to the bath modes at a given frequency $\omega'$. The key simplification, known as the Markov approximation, assumes that the bath's memory is very short and that the coupling strength $\kappa(\omega')$ is essentially constant over the range of frequencies relevant to the system's dynamics.

Therefore, we can approximate $\kappa(\omega')$ by its value at the central frequency of interest, $\omega_c$, and treat it as a constant. This constant, $\kappa_c$, represents the effective decay rate of the system into the bath:
\begin{equation}\label{eq:kappa_def_revised}
\kappa(\omega') \approx \kappa(\omega_c) \equiv \kappa_c.
\end{equation}

In the given context, $\kappa$ represents the \textit{rate of transfer of excitations}, akin to Fermi's golden rule. It is used in the approximation where the summation term of Eq. (\ref{eq:kappa_def}) is approximated as $\kappa(\omega_c + \omega)$, where $\omega_c$ is a \textit{central frequency} resonating with the free evolution of $\hat{a}(t)$. This approximation simplifies the calculation and allows for the interpretation of the term as the rate of transfer of excitations.

Another valuable approximation assumes the coupling strength coefficients $g_n$ to be approximately a real constant $g$. This assumption is reasonable given the frequencies of interest $\omega_c + \omega \approx \omega_c$. This approximation simplifies the expression for $\kappa_c$ by assuming that the coupling strength $g$ is roughly constant across the range of frequencies relevant to the system's dynamics. 

The condition $\omega_c + \omega \approx \omega_c$ is a central feature of the Markov approximation. It is justified by analyzing the system in a reference frame rotating at the central frequency $\omega_c$. In this frame, the system's operators evolve slowly. A slowly varying function is, by its nature, dominated by low-frequency Fourier components. This means the only significant contributions to the interaction integral come from the Fourier frequency variable $\omega$ being close to zero. For these dominant components, approximating the total frequency $\omega_c + \omega$ by $\omega_c$ becomes a good and physically motivated simplification. Consequently: 
\begin{equation}
\kappa_c = 2\pi \sum_{n} |g_n|^2 \delta(\omega_c - \omega_n) 
 \approx g^2 \sum_{n} \delta(\omega_c + \omega_n) = 2\pi g^2 D(\omega_c).
 \end{equation}

The expression for $\kappa_c$ encapsulates the effective coupling rate between a quantum system and its surrounding bath of harmonic oscillators. From a microscopic perspective, $\kappa_c$ is defined as a discrete summation over bath modes weighted by their respective coupling strengths $|g_n|^2$ and constrained by the energy-conserving Dirac delta function $\delta(\omega_c - \omega_n)$, ensuring that only resonant interactions contribute to the energy exchange.

Under the assumption that the coupling coefficients $g_n$ vary slowly with frequency and can be approximated by a constant $g$, the summation simplifies by substituting $|g_n|^2 \approx g^2$. Moreover, the Dirac delta function enforces resonance and effectively selects modes at frequency $\omega_c$. The summation $\sum_n \delta(\omega_c - \omega_n)$ then defines the spectral density of the bath, $D(\omega_c)$, which quantifies the number of available bath modes at the system frequency. 

As a result, the expression reduces to $\kappa_c = 2\pi g^2 D(\omega_c)$, establishing a direct relationship between the effective decay rate, the coupling strength, and the bath's density of states at the resonance frequency. This approximation underpins the Markovian treatment of system-bath interactions and is central to the derivation of input-output theory and Langevin dynamics in quantum optics and circuit QED.
  
Reverting to Eq. (\ref{eq:drives_kernel_diracs}) and assuming $\omega_n - \omega_c = \omega$, it is obtained:
\begin{equation} \label{eq:drives_kernel_approx} 
K(t - t') = \int_{-\infty}^{\infty} \frac{d\omega}{2\pi} \kappa(\omega_c + \omega) e^{-i\omega(t - t')} 
\approx \kappa_c \delta(t - t').
\end{equation}
Employing Eq. (\ref{eq:drives_kernel_approx}) in Eq. (\ref{eq:drives_approx_reform}) yields:
\begin{equation}
    \int_{0}^{t} dt' \hat{a}(t') e^{-i\omega_c(t - t')} \kappa_c \delta(t-t') = \frac{\kappa_c}{2} \hat{a},
\end{equation}
integrating these approximations into Eq. (\ref{eq:motion_a_bath}):
\begin{equation}\frac{d}{dt}\hat{a} = \frac{i}{\hbar}[\hat{H}_s, \hat{a}] 
-i\sqrt{\frac{\kappa_c}{2\pi D(\omega_c)}}\sum_n \hat{b}_{n}^0 e^{-i \omega_n t} - \frac{\kappa_c}{2} \hat{a},
\end{equation}
furthermore, by redefining:
\begin{equation}\label{eq:drives_bin_def}
    \hat{b}_{in}(t) = \frac{i}{\sqrt{2\pi D(\omega_c)}} \sum_{n} \hat{b}_{n}^0 e^{-i\omega_n t},
\end{equation}
it is possible to arrive at the known Langevin equation:
\begin{equation}\label{eq:drives_langevin}
  \frac{d}{dt}\hat{a} = \frac{i}{\hbar}[\hat{H}_s, \hat{a}] - \frac{\kappa_c}{2}\hat{a} - \sqrt{\kappa_c} \hat{b}_{in},  
\end{equation}
where $\hat{b}_{in}$ represents an \textit{external drive acting upon the system of interest}. It manifests as the Fourier transform of the harmonic amplitudes $\hat{b}_{n}^0$, and is hence proportional to an applied time-dependent signal. Notably, $\hat{b}_{in}$ remains unaffected by $\hat{a}$ but influences the cavity modes as delineated in the Langevin equation.

If instead of Eq. (\ref{eq:drives_bn_def}), it is shown a solution for $\hat{b}_n(t)$ in terms of a final time $t_f > t$:
\begin{equation}
\hat{b}_n(t) = \hat{b}_{n}^f e^{-i\omega_n(t-t_f)} 
+ig_n^* \int_{t}^{t_f} dt' \hat{a}(t') e^{-i\omega_n(t-t')}.
\end{equation}

Hence, an alternative form of the Langevin equation:
\begin{equation}
    \frac{d}{dt}\hat{a} = \frac{i}{\hbar}[\hat{H}_s, \hat{a}] +\frac{\kappa_c}{2}\hat{a} -  \sqrt{\kappa_c} \hat{b}_{out},
\end{equation}
where
\begin{equation}
    \hat{b}_{out}(t) = \frac{i}{\sqrt{2\pi D(\omega_c)}} \sum_{n} \hat{b}_{n}^f e^{-i\omega_n(t-t_f)}.
\end{equation}

By subtracting both forms of the Langevin equation, we obtain the input/output relation:
\begin{equation}
    \hat{b}_{out}(t) = \hat{b}_{in}(t) + \sqrt{\kappa_c} \hat{a}(t).
\end{equation}

While this formulation is general, it can be interpreted in the context of physical systems. If a coupled transmission line simulates the thermal bath, both $\hat{b}_{in}$ and $\hat{b}_{out}$ signify time-dependent signals propagating through the line towards and away from the system of interest. For instance, if the transmission line is defined in the region $0 < x < L$ and linked to a system at $x = 0$, the components $\hat{b}_{in}^0$ correspond to the operators of left-traveling waves $\hat{b}_L$ as defined earlier. Analogously, the components $\hat{b}_{out}^0$ refer to the right-traveling modes $\hat{b}_R$.

Now we will see what happens when a resonator is coupled with a thermal bath. The Hamiltonian that describes the resonator $\hat{H}_s = \hbar\omega_r \hat{a}^\dagger\hat{a}$. Eq. (\ref{eq:drives_langevin}) then becomes:
\begin{equation}
    \frac{d}{dt}\hat{a} = -i\omega_r\hat{a} - \frac{\kappa_c}{2}\hat{a} - \sqrt{\kappa_c}\hat{b}_{in}.
\end{equation}

This set of linear differential equations can be solved using the Laplace transform to yield:
\begin{equation}
\hat{a}[s] = \frac{1}{s + i\omega_r + \kappa_c/2}\hat{a}(0) 
-\frac{\sqrt{\kappa_c}}{s + i\omega_r + \kappa_c/2}\hat{b}_{in}[s],
\end{equation}
where $\hat{a}[s]$ and $\hat{b}_{in}[s]$ represent the \textit{s-plane components of the respective time-dependent operators}. 

The Laplace transform is employed instead of the Fourier transform due to its ability to incorporate initial conditions and capture transient dynamics in causal systems. This is useful for modeling open quantum systems, where decay and relaxation processes dominate. Moreover, it facilitates input-output analysis and stability characterization in the complex frequency domain, making it a natural choice for describing dissipative quantum dynamics. For $\hat{b}_{out}$:
\begin{equation}
\hat{b}_{out}[s] = \frac{\sqrt{\kappa_c}}{s + i\omega_r + \kappa_c/2}\hat{a}(0) 
+ \frac{s + i\omega_r - \kappa_c/2}{s + i\omega_r + \kappa_c/2}\hat{b}_{in}[s].
\end{equation}

The reflection coefficient $R(s)$ of the resonator derived from this equation is, as $\bra\hat{a}(0) \ket = 0$:
\begin{equation}\label{eq:resonator_reflection}
    R(s) = \frac{\bra\hat{b}_{out}[s] \ket}{\bra\hat{b}_{in}[s] \ket} = \frac{s + i\omega_r - \kappa_c/2}{s + i\omega_r + \kappa_c/2}.
\end{equation}

In the expression for $R(s)$, the numerator and the denominator contain the complex frequency $s = \sigma + i\omega$, which accounts for the dynamics of the system. The imaginary part, $\omega$, represents the frequency of the probe signal, while the real part, $\sigma$, represents its exponential decay or growth rate. The term $\omega_r$ is the natural resonant frequency of the cavity, and the parameter $\kappa_c/2$ is the damping rate of the resonator due to its coupling to the environment.

Moreover, this coefficient characterizes how the resonator responds to an input signal $\hat{b}_{in}[s]$ by relating the output signal $\hat{b}_{out}[s]$ to the input signal in the Laplace domain. It delineates the reflection coefficient's reliance on the resonant frequency of the cavity. This frequency becomes modifiable when the cavity establishes a dispersive coupling with a transmon, contingent upon the transmon's state. Consequently, the assessment of $R(s)$ proves instrumental in discerning the populated or unpopulated state of the transmon. The main technique to perform this distinction is via dispersive readout, which will be discussed further.

\subsubsection{Dispersive Readout}
In the realm of cQED, dispersive readout stands as a crucial technique for probing and manipulating quantum systems. This approach enables the extraction of valuable information about the state of a quantum system by observing the frequency shift induced by its interaction with a measurement apparatus. 

To model the measurement process, we consider an input signal consisting of a single, coherent microwave tone at a specific probe frequency, $\omega_s$. In the frequency domain, this input field, $\hat{b}_{in}(\omega)$, can be represented as:
\begin{equation}
\hat{b}_{in}(\omega) = \beta_s \delta(\omega - \omega_s),
\end{equation}
where, $\hat{b}_{in}(\omega)$ is the frequency-domain representation of the input field operator previously introduced as $\hat{b}_{in}(t)$. The Dirac delta function, $\delta(\omega - \omega_s)$, enforces the condition that the input field only contains excitations at the single frequency $\omega_s$. $\beta_s$ is the classical complex amplitude of the coherent input drive. Its magnitude squared, $|\beta_s|^2$, is proportional to the input power of the microwave tone. Using a classical amplitude instead of an operator is justified because a coherent state is the quantum state that most closely resembles a classical field.

Here, $\hat{b}_{in}(\omega)$ is the frequency-domain representation of the input field operator, the Dirac delta function $\delta(\omega - \omega_s)$ enforces the single-frequency condition, and $\beta_s$ is the classical complex amplitude of the coherent input drive, whose magnitude squared, $|\beta_s|^2$, is proportional to the input power.

The complex voltage conveyed by the transmission line to the resonator's location at $x = 0$ is derived from the positive-frequency part of the voltage operator, $\hat{V}^{(+)}$. Starting from the general expression in Eq. (\ref{eq:voltage_expression}) and substituting the single-tone input field yields:
\begin{align}
    \hat{V}^{(+)}(0, t) &= -i\int_{0}^{\infty} d\omega \sqrt{\frac{\hbar \omega}{4\pi c_0 v}} \hat{b}_{in}(\omega) e^{-i\omega t} \\
    &= -i \sqrt{\frac{\hbar \omega_s}{4\pi c_0 v}} \beta_s e^{-i\omega_s t}.
\end{align}
This expression gives the complex amplitude of the voltage signal at frequency $\omega_s$ that arrives at the resonator. The physically meaningful relation connects the power of the input signal to this voltage. The average input power $P_{in}$ is given by the photon rate multiplied by the energy per photon, $P_{in} = \hbar\omega_s |\beta_s|^2$. In a transmission line, power is also $P_{in} = V_{\text{rms}}^2 / Z_0$, where $Z_0 = 1/(v_0 c_0)$ is the characteristic impedance. Equating this gives the fundamental relationship between voltage and the quantum field amplitude:
\begin{equation} \label{eq:drives_vb_ratio_corrected}
    V_{\text{rms}} = \sqrt{\hbar \omega_s Z_0} |\beta_s|.
\end{equation}

This frequency-dependent relation implies that $\hat{V}$ is not directly proportional to $\hat{b}_{in}$ when the input signal comprises a broad spectrum of frequency components. In such cases, each frequency contributes differently to the voltage due to the $\sqrt{\omega}$ dependence, preventing a uniform scaling between $\hat{b}_{in}(t)$ and $\hat{V}(t)$. Only in the narrowband limit, where the frequency content is sharply peaked around a central frequency $\omega_s$, can the voltage be considered approximately proportional to the input amplitude.

In the dispersive regime of resonator-transmon coupling, the resonator's resonant frequency shifts depending on the state of the transmon. This occurs when the qubit (at frequency $\omega_q$) and resonator (at frequency $\omega_r$) are far detuned, i.e., $|\Delta| = |\omega_q - \omega_r| \gg g$, where $g$ is their coupling strength. In this limit, they cannot directly exchange energy, but instead cause a state-dependent frequency shift $\chi \approx g^2/\Delta$ \cite{Blais2004, Wallraff2005_dispersive}. This modification alters the Langevin equation, which must be conditioned on the qubit state ($g$ for ground, $e$ for excited):
\begin{equation}
    \frac{d}{dt}\hat{a}_{g,e} = -i\left(\omega_r \mp \frac{\chi}{2}\right)\hat{a}_{g,e} -\frac{\kappa_c}{2}\hat{a}_{g,e} - \sqrt{\kappa_c}\hat{b}_{in}.
\end{equation}
Upon replacing $\hat{a}_{g,e}$ with $e^{-i\omega_r t}\tilde{a}_{g,e}$, there is a shift to a reference frame where the cavity states exhibit slow rotation:
\begin{equation}
    \frac{d}{dt}\tilde{a}_{g,e} = \left(\mp i\frac{\chi}{2} - \frac{\kappa_c}{2}\right)\tilde{a}_{g,e} - \sqrt{\kappa_c}\tilde{b}_{in}(t),
\end{equation}
Here, $\tilde{b}_{in}(t) = \hat{b}_{in}(t) e^{i\omega_r t}$ now correlates with the baseband signal transmitted to the cavity. 

By exploiting the frequency shift induced by the interaction between a quantum system and a measurement apparatus, dispersive readout allows for the extraction of important information about the system's state. In other words, instead of directly measuring certain properties of the quantum system, such as its energy or amplitude, dispersive readout relies on detecting changes in the system's resonant frequency due to its interaction with the measurement apparatus. This approach enables researchers to extract important details about the quantum system's state without directly perturbing it, making it a valuable tool in quantum information processing and quantum computing applications.

\subsubsection{Displaced Frame}
The concept of a displaced frame offers an alternative approach, steering away from the intricate quantum treatment of an external drive, particularly necessary for input/output theory. A classical drive is typically implemented by coupling a voltage source to the resonator. This interaction is described by a Hamiltonian term $\hat{H}_{\text{drive}} = -\hat{Q}\hat{V}(t)$, where $\hat{Q}$ is the resonator's charge operator and $\hat{V}(t)$ is the classical input voltage. The charge is given by $\hat{Q} \propto i(\hat{a}^\dagger - \hat{a})$, and for a sinusoidal drive at frequency $\omega_d$, the voltage is $V(t) \propto \cos(\omega_d t)$. The interaction term thus contains products like $\hat{a}^\dagger e^{i\omega_d t}$ and $\hat{a} e^{-i\omega_d t}$. By applying the RWA to neglect the fast-oscillating, counter-rotating terms, the total time-dependent Hamiltonian for the driven resonator takes the form:
\begin{equation}
    \hat{H}(t) = \hbar \omega_r \hat{a}^\dagger \hat{a} + \hbar (\epsilon_d e^{-i\omega_d t} \hat{a}^\dagger + \epsilon_d^* e^{i\omega_d t} \hat{a}),
\end{equation}
where $\omega_r$ is the resonator frequency, $\omega_d$ is the drive frequency, and $\epsilon_d$ is the complex amplitude of the drive, encompassing both its strength and phase.

To eliminate the explicit time dependence, we transition to a rotating frame defined by the unitary transformation:
\begin{equation}
    \hat{U}(t) = e^{i\omega_d t \hat{a}^\dagger \hat{a}}.
\end{equation}
The transformed Hamiltonian in this interaction picture is given by:
\begin{equation}
    \hat{H}' = \hat{U} \hat{H} \hat{U}^\dagger - i\hbar \hat{U} \frac{d}{dt} \hat{U}^\dagger.
\end{equation}
Evaluating each term yields:
\begin{align}
    \hat{U} \hat{H} \hat{U}^\dagger &= \hbar \omega_r \hat{a}^\dagger \hat{a} + \hbar (\epsilon_d \hat{a}^\dagger + \epsilon_d^* \hat{a}), \\
    -i\hbar \hat{U} \frac{d}{dt} \hat{U}^\dagger &= -\hbar \omega_d \hat{a}^\dagger \hat{a}.
\end{align}
Thus, the effective Hamiltonian in the rotating frame becomes:
\begin{equation}
    \hat{H}_{\text{eff}} = \hbar (\omega_r - \omega_d) \hat{a}^\dagger \hat{a} + \hbar (\epsilon_d \hat{a}^\dagger + \epsilon_d^* \hat{a}).
\end{equation}
This is referred to as an “effective” Hamiltonian because it accurately describes the system's dynamics in the rotating frame, with the fast oscillations at the drive frequency $\omega_d$ having been mathematically removed. Let us define the detuning as $\Delta = \omega_r - \omega_d$. The Hamiltonian is a time-independent operator that correctly captures the drive dynamics:
\begin{equation}\label{eq:displaced_frame_hamiltonian}
    \hat{H}_{\text{eff}} = \hbar\Delta \hat{a}^\dagger \hat{a} + \hbar (\epsilon_d\hat{a}^\dagger + \epsilon_d^*\hat{a}).
\end{equation} 

The evolution of the annihilation operator in this frame follows the Heisenberg equation of motion~\cite{sakurai_napolitano_2020}:
\begin{equation}\label{eq:heisenberg_eq_motion}
    \frac{d}{dt}\hat{a} = \frac{1}{i\hbar}[\hat{H}_{\text{eff}}, \hat{a}] = \frac{1}{i\hbar} \left( \hbar\Delta[\hat{a}^\dagger \hat{a}, \hat{a}] + \hbar\epsilon_d[\hat{a}^\dagger, \hat{a}] \right) = -i\Delta \hat{a} - i\epsilon_d.
\end{equation}

Within this framework, incorporating losses can be approached in two ways: formulating a Lindblad master equation~\cite{Manzano2019A} encompassing amplitude damping, or simply appending a phenomenological damping term to the Heisenberg equation:
\begin{equation}
    \frac{d}{dt}\hat{a} = -i\Delta \hat{a} - \frac{\kappa}{2} \hat{a} - i\epsilon_d.
\end{equation}
Here, $\kappa$ represents the \textit{decay constant}, measured in excitations per unit time.

It is important to note that, while both the phenomenological inclusion of a damping term in the Heisenberg equation and the Lindblad master equation approach aim to describe dissipation, they are not strictly equivalent. The Lindblad formalism stems from a rigorous treatment of open quantum systems under the Born-Markov and secular approximations, ensuring complete positivity and trace preservation of the density matrix. In contrast, appending a decay term directly to the Heisenberg equation is a heuristic method that, although often accurate in the semiclassical or weak-damping regimes, may fail to capture certain quantum statistical correlations and non-Markovian effects present in a fully quantum description.

The essence of the displaced frame resides in absorbing the drive $\epsilon_d$ into the definition of $\hat{a}$. Introducing a classical parameter $\xi_d$, which obeys the same equation of motion as the expectation value of $\langle\hat{a}\rangle$:
\begin{equation}
    \frac{d}{dt} \xi_d = -i\Delta \xi_d - \frac{\kappa}{2} \xi_d - i\epsilon_d.
\end{equation}

The difference between the last two equations yields an equation for a new operator, $\tilde{a} = \hat{a} - \xi_d$:
\begin{equation}
    \frac{d}{dt}\tilde{a} = -i\Delta \tilde{a} - \frac{\kappa}{2} \tilde{a}.
\end{equation}
Notably, in this new “displaced” frame, the drive does not directly influence the mode operators but rather displaces the frame itself. Correspondingly, the associated unitary operation constitutes a displacement:
\begin{equation}
    \hat{U}_d(t) = e^{\xi_d(t) \hat{a}^\dagger - \xi_d^*(t) \hat{a}},
\end{equation}
whereby, $\hat{U}_d^\dagger \hat{a} \hat{U}_d = \hat{a} - \xi_d = \tilde{a}$. Assuming the drive manifests as a constant pulse at frequency $\omega_d$, the steady-state solution for the displacement parameter is derived as:
\begin{equation}
    \xi_d = \frac{-i\epsilon_d}{\kappa/2 + i(\omega_r - \omega_d)}.
\end{equation}

This last equation shows the steady-state complex amplitude of the displacement, $\xi_d$, is determined by the drive amplitude ($\epsilon_d$), the system's decay rate ($\kappa$), and the detuning between the drive and resonator frequencies ($\omega_d - \omega_r$). Overall, this equation describes the coherent state amplitude of the resonator field in response to the applied drive.

\section{Application Example: Rabi Oscillations in a Driven Transmon-Resonator System}\label{sec:app}

This section consolidates the preceding theoretical concepts by presenting a numerical simulation of a canonical experiment in circuit QED: the observation of vacuum Rabi oscillations. This phenomenon, which represents the coherent exchange of a single quantum of energy between a two-level atom and a cavity mode, serves as a fundamental signature of the strong coupling regime. The simulation will model a driven transmon qubit coupled to a microwave resonator, incorporating the Jaynes-Cummings Hamiltonian, external drives, and the effects of environmental dissipation. The numerical solution of the system's dynamics will be performed using the \texttt{QuTiP} (Quantum Toolbox in Python) library~\cite{qutip1, qutip2}, a standard tool for simulating open quantum systems.

\subsection{System Configuration and Parameters}

The physical system consists of a transmon qubit coupled to a high-quality factor cavity. Table~\ref{tab:params} summarizes the simulation parameters.

\begin{table}[h]
\centering
\caption{Simulation parameters for the transmon-resonator system.}
\label{tab:params}
\begin{tabular}{ll}
\hline
\textbf{Parameter} & \textbf{Value} \\
\hline
Cavity dimension (Fock basis) & $N = 10$ \\
Cavity frequency $\omega_R/2\pi$ & $7.0$ GHz \\
Qubit frequency $\omega_T/2\pi$ & $5.0$ GHz \\
Coupling strength $g/2\pi$ & $200$ MHz \\
Drive amplitude $A/2\pi$ & $160$ MHz \\
Time resolution & $256$ points \\
\hline
\end{tabular}
\end{table}

The system is initialized in the ground state, with the cavity and qubit frequencies initially far detuned to suppress resonant interaction. Detuning variation is later introduced to explore resonant and near-resonant dynamics.

\subsection{Experimental Protocol and Simulation Scenarios}

The simulated system consists of a two-level superconducting qubit coupled to a single-mode resonator, modeled via the Jaynes–Cummings Hamiltonian. The protocol explores dynamical regimes by varying coupling strength, dissipation, and detuning, enabling the investigation of coherent and open-system behavior.

The following scenarios are considered:

\begin{enumerate}
    \item \textbf{Baseline System Behavior}: Uncoupled dynamics (\(g = 0\)) Establishes a baseline by isolating the qubit and cavity evolutions. The addition of the coupling factor adds an energy exchange behavior. Closed versus open dynamics compare unitary evolution to dissipative scenarios incorporating relaxation and photon loss through collapse operators.
    
    \item \textbf{Vacuum Rabi Oscillations Measurement}: The energy exchange between qubit, cavity and applied drive is evaluated in terms of a frequency detuning between the qubit and the cavity.
\end{enumerate}

State populations are tracked over time, revealing coherent oscillations, vacuum Rabi splitting, and decoherence mechanisms across different physical regimes.

\subsection{Hamiltonian and Dissipative Dynamics}

The coherent evolution of the system is governed by the Jaynes-Cummings Hamiltonian under the RWA. To model the irreversible effects of environmental interaction, such as dissipation and decoherence, the system's evolution is described not by the state vector, but by the density matrix $\rho$. The dynamics of the density matrix are governed by the Lindblad master equation:
\begin{equation}
    \frac{d\rho}{dt} = \frac{1}{i\hbar}[\hat{H}, \rho] + \sum_{k} \mathcal{D}[L_k]\rho,
\end{equation}
where the first term describes the unitary evolution and the second term, the dissipator, describes the effects of the environment. Each dissipation channel is represented by a \textbf{Lindblad collapse operator}, $L_k$, where $\mathcal{D}[L_k]\rho = L_k \rho L_k^\dagger - \frac{1}{2}\{L_k^\dagger L_k, \rho\}$.

For the transmon-resonator system, the primary dissipative dynamics include photon loss from the cavity and relaxation of the qubit. These are described by the following collapse operators:
\begin{subequations}\label{eq:collapse_oper}
\begin{align}
    L_1 &= \sqrt{\kappa(1+\langle n_{th} \rangle)}\, \hat{a}; \\
    L_2 &= \sqrt{\kappa \langle n_{th} \rangle}\, \hat{a}^\dagger; \\
    L_3 &= \sqrt{\gamma}\, \hat{\sigma}_-.
\end{align}
\end{subequations}

Here, $\kappa$ is the cavity photon decay rate, $\gamma$ is the qubit energy relaxation rate, and $\langle n_{th} \rangle$ is the average number of thermal photons in the cavity bath. In the zero-temperature limit ($\langle n_{th} \rangle \to 0$), the operator $L_2$ vanishes and only energy decay processes are considered. The presence of $L_2$ in the finite-temperature case models the absorption of thermal photons from the environment.

\subsection{Measurement}

After evolution under each configuration, the occupation probabilities of the transmon and resonator are extracted from the density matrix. These results reveal coherent dynamics, such as vacuum Rabi oscillations, and quantify the impact of detuning and dissipation.

\subsection{Transmon Excitation via Coherent Drive}

Excitation of the transmon qubit is modeled through an external coherent drive field, which couples to the qubit's dipole moment and induces Rabi oscillations. This approach reflects experimental protocols in circuit QED, where microwave tones are applied to the system to manipulate its quantum state.

The drive is incorporated into the system Hamiltonian as a time-dependent term of the form:

\begin{equation}
    \hat{H}_{\text{drive}} = A \left(e^{-i\omega_d t}\hat{b}^\dagger + e^{i\omega_d t}\hat{b} \right),
    \label{eq:ham_drive}
\end{equation}
where $A$ is the drive amplitude, $\omega_d$ is the drive frequency, and $\hat{b}$ ($\hat{b}^\dagger$) is the annihilation (creation) operator acting on the transmon qubit's Hilbert space, treated here as a two-level system. This Hamiltonian describes the interaction of the transmon with a classical field oscillating at frequency $\omega_d$.

Qualitatively, the drive term enables coherent transitions between the ground and excited states of the qubit. When $\omega_d$ is resonant or near-resonant with the transmon frequency $\omega_T$, the drive can efficiently induce Rabi oscillations, allowing precise control over the excitation amplitude and duration. The drive amplitude $A$ determines the Rabi frequency, setting the rate at which the qubit undergoes state transitions.

In the simulation protocol, the transmon is initially decoupled from the cavity and driven by the external field until it reaches the desired excited-state population. Following this preparation step, the coupling between the transmon and resonator is activated to investigate energy exchange and coherent dynamics.

\subsection{Simplified Experimental Setup}
The simulation proceeds as a standard experimental setup for cQED, directly reflecting the theoretical concepts discussed. At room temperature, an Arbitrary Waveform Generator (AWG) produces precisely shaped baseband pulses, which are mixed with a stable microwave carrier from a Local Oscillator (LO) using an IQ Mixer. This process generates the high-frequency “Drive Signal” used to manipulate the qubit's state, as described by the drive Hamiltonian. This signal is heavily attenuated and sent down into the Cryogenic Stage, which operates at millikelvin temperatures to prevent thermal noise from destroying the fragile quantum states. Inside, the drive signal interacts with the Qubit-Resonator System. To measure the qubit's state, a separate “Readout Signal” is sent to the resonator. The signal interacts with the system and is reflected, carrying information about the qubit's state via the dispersive shift. This faint reflected signal is then routed back to room temperature, where it is amplified and demodulated by the Readout Electronics to determine the final qubit state, completing the measurement process.

\begin{figure}[!ht]
    \centering
    \includegraphics[width = 76mm]{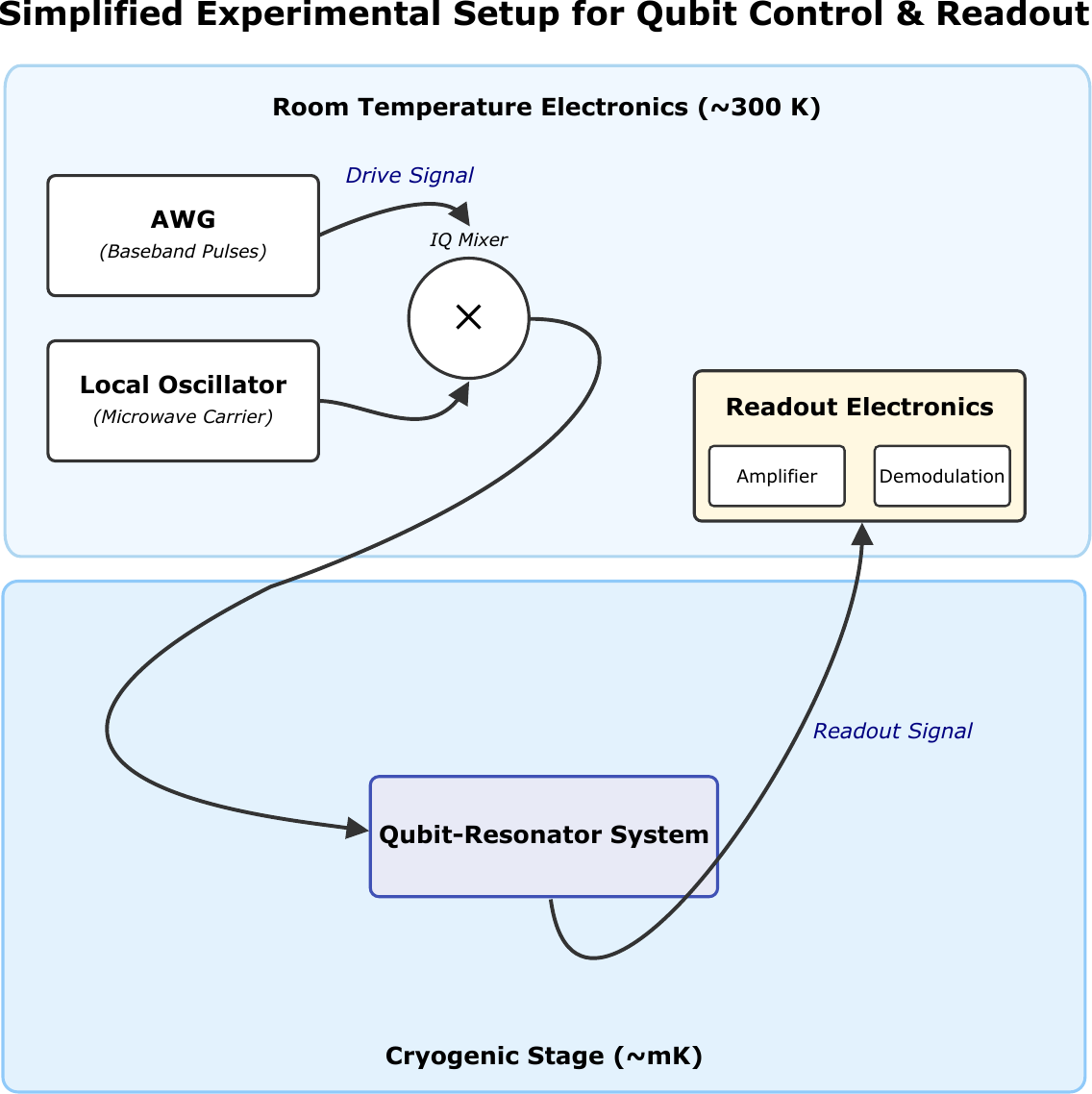}
    \caption{A simplified schematic of a typical experimental setup for superconducting qubit control and readout. Room-temperature electronics generate and shape microwave pulses (Drive Signal) that are sent into a cryogenic environment to manipulate the state of the Qubit-Resonator System. A separate Readout Signal probes the resonator, and the reflected signal is amplified and processed to measure the qubit's state.}
    \label{fig:setupSchema}
\end{figure}

\subsection{System Evolution — Results and Discussion}
In this section, different configurations of the aforementioned system are simulated, and the results are shown with their respective discussions.

\subsubsection{System baseline behavior}
Firstly, it is important to observe the interaction between the qubit and the resonator, which is governed by a coherent coupling term of strength \(g\), without the external drive applied. This interaction occurs through the exchange of one photon, initially confined within the cavity. Dissipative dynamics are incorporated using collapse operators that model energy relaxation in both subsystems. 

\begin{figure}[!ht]
    \centering
    \includegraphics[width = 76mm]{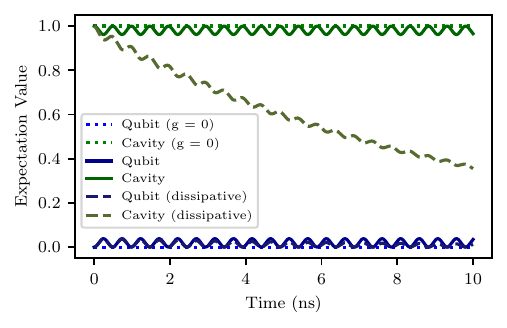}
    \caption{Expectation values of the qubit and cavity occupation numbers under various conditions. Solid lines show coherent dynamics with qubit–cavity coupling $g \neq 0$; dotted lines indicate the uncoupled case $g = 0$; and dashed lines represent dynamics in the presence of dissipation. The cavity maintains high excitation, while dissipation causes a gradual energy loss, particularly in the cavity mode. The qubit remains weakly excited, confirming that the drive predominantly affects the cavity.}
    \label{fig:simulation_noDrive}
\end{figure}

The baseline behavior is illustrated in Fig.~\ref{fig:simulation_noDrive}.
In the absence of coupling (\(g = 0\)), the qubit and cavity evolve independently. The cavity maintains its population in the absence of any excitation or decay mechanisms. Once the coherent coupling is introduced, periodic energy exchange between the qubit and the resonator is observed, characteristic of Rabi oscillations in the Jaynes–Cummings regime. The inclusion of dissipation, modeled via a Lindblad master equation, results in a monotonic decay of the cavity excitation, indicating photon leakage from the resonator. 

Notably, the qubit remains near its ground state throughout the evolution, suggesting that no external energy is inserted into the system and the energy exchange with the cavity is not sufficient to make the qubit reach higher expectation values under the parameters considered. This dynamic highlights the importance of spectral matching and coupling topology in engineering coherent control and dissipation pathways in hybrid quantum systems.

\subsubsection{Vacuum Rabi Oscillations Measurement}
To simulate the vacuum Rabi dynamics of a two-level superconducting qubit coupled to a single-mode quantized electromagnetic field inside a resonator, the Hamiltonian governing the evolution includes the drive coupling mechanism seen in Eq.~(\ref{eq:ham_drive}). 

\begin{figure}[!ht]
    \centering
    \includegraphics[width = 76mm]{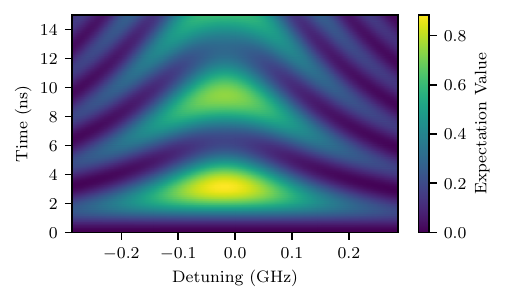}
    \caption{Vacuum Rabi oscillations as a function of qubit–cavity detuning.
        Color map of the excited state population of the qubit as a function of interaction time and frequency detuning between the qubit and the resonator. The data reveal the characteristic vacuum Rabi splitting and oscillatory exchange of energy between the qubit and the resonator, with maximum excitation probability occurring at resonance (\(\Delta = 0\)).}
    \label{fig:3DRabi}
\end{figure}

The system is initialized with the qubit in its excited state and the cavity in its vacuum state, representing a single excitation in the combined Hilbert space. The time evolution is computed for a range of detunings \(\Delta\), and the excited-state population of the qubit is recorded as a function of both time and detuning. The resulting interference pattern exhibits characteristic chevron-shaped oscillations centered at zero detuning, corresponding to resonant energy exchange between the qubit and the cavity mode. The off-resonant cases show slower and less complete oscillations due to energy mismatch. The result can be visualized in Fig.~\ref{fig:3DRabi}.

This protocol effectively reveals the coherent nature of light–matter interaction in the cQED regime, where strong coupling allows for reversible exchange of excitations between the matter-like and photonic degrees of freedom. The observed vacuum Rabi splitting serves as a signature of coherent hybridization, and the dynamics are indicative of fundamental quantum behavior in engineered light–matter systems.

\section{Conclusion}

This tutorial established a foundation in the principles and applications of superconducting quantum circuits, progressing from fundamental phenomena to advanced quantum electrodynamic architectures. The initial sections elucidated the macroscopic quantum behavior of superconductors, including the Meissner effect and flux quantization, employing both classical and quantum mechanical frameworks. The London equations were derived as essential descriptors of ideal conductivity and perfect diamagnetism, forming the theoretical basis for subsequent quantum device discussions.

A quantum mechanical treatment of the Josephson effect revealed the implications of phase coherence across weak links, with the DC and AC Josephson relations formalizing the interplay between supercurrents, phase differences, and electromagnetic potentials. This framework extended to cQED, detailing the quantization of electromagnetic modes in lumped-element and transmission-line resonators. The formalism demonstrated the function of superconducting resonators as quantum buses for coherent information transfer and infrastructure for qubit coupling.

The transmon qubit was presented as a foundational platform, with its design principles and operational regimes analyzed. By utilizing the nonlinearity of Josephson junctions and mitigating charge noise via large \(E_J/E_C\) ratios, transmons achieve enhanced coherence times critical for quantum information processing. The derivation of the transmon-resonator coupling Hamiltonian, complemented by the rotating wave approximation and Jaynes-Cummings model, elucidated mechanisms of coherent energy exchange and state hybridization. Practical considerations, including drive engineering, dissipative dynamics through input-output theory, and dispersive readout techniques, were integrated to connect theoretical constructs with experimental implementations.

Numerical simulation of vacuum Rabi oscillations in a driven transmon-resonator system demonstrated the practical application of these principles. Parameter sweeps across detuning, coupling strength, and dissipation regimes visualized phenomena such as state oscillations and chevron interference patterns, illustrating the sensitivity of quantum dynamics to Hamiltonian parameters and environmental interactions. This case study reinforced cQED's capability to emulate and probe light-matter interactions at the quantum level.

The principles and applications of superconducting quantum circuits, as presented, position them as a versatile and scalable platform for quantum information science and technology. The systematic exposition, from macroscopic quantum phenomena to cQED architectures, provides a foundational framework for research and development. The demonstrated capacity for coherent quantum control, along with insights from numerical simulations of light-matter interactions, supports the potential of these engineered quantum systems.

\section*{Funding}
The authors received no specific funding for this work.

\addcontentsline{toc}{section}{References} 
\section*{References}
\nocite{*}
\bibliographystyle{unsrt}
\bibliography{sn-bibliography}

\appendix

\section{Quantum Harmonic Oscillator~\cite{sakurai_napolitano_2020}}\label{secA1}

\subsubsection{Energy and Important Operators}
The simple harmonic oscillator is described by its basic Hamiltonian:
\begin{equation}\label{eq:class_harm_osc}
    \hat{H} = \frac{\hat{p}^2}{2m} + \frac{m \omega^2 \hat{x}^2}{2},
\end{equation}
where $\omega$ is the angular frequency of the classical oscillator related to the spring constant $k$ in Hooke's law via $\omega = \sqrt{\frac{k}{m}}$. The operators $\hat{x}$ and $\hat{p}$ are Hermitian. Moreover, the Hamiltonian can be factored as:
\[\hat{H} = \hbar \omega \left( \frac{m\omega}{2\hbar} \right) \times\]  
 \begin{equation}
\times \left[\left(\hat{x} + \frac{i\hat{p}}{m\omega} \right)\left(\hat{x} - \frac{i\hat{p}}{m\omega} \right) + \frac{1}{2} \right].
\end{equation}

Furthermore, it is convenient to define two non-Hermitian operators known as the annihilation operator $\hat{a}$ and the creation operator $\hat{a}^\dagger$:
\begin{equation}
\left\{
\begin{array}{cc}
    \hat{a} &= \sqrt{\dfrac{m\omega}{2\hbar}} \left(\hat{x} + \dfrac{i\hat{p}}{m\omega}\right)   \\ & \\
    \hat{a}^\dagger &= \sqrt{\dfrac{m\omega}{2\hbar}} \left(\hat{x} - \dfrac{i\hat{p}}{m\omega}\right) 
\end{array}
\right. .
\end{equation}

Using the canonical commutation relations, it is obtained:
\begin{equation}\label{eq:number_commutation}
    [\hat{a}, \hat{a}^\dagger] = 1.
\end{equation}

Given these relations, it is defined the number operator $\hat{N}$:
\begin{equation}
    \hat{N} = \hat{a}^\dagger \hat{a},
\end{equation}
which is Hermitian. With that, the Hamiltonian can be written as:
\begin{equation}
    \hat{H} = \hbar \omega \left(\hat{N} + \frac{1}{2} \right).
\end{equation}

That is an important relation between the number operator and the Hamiltonian operator. Since the Hamiltonian $\hat{H}$ is a straightforward function of the number operator $\hat{N}$, it is possible to, simultaneously, find the eigenvalues of $\hat{N}$ along with those of $\hat{H}$. The representation of an energy eigenstate of $\hat{N}$ by the value $n$, such that $\hat{N}|n\ket = n|n\ket$, so that $n$ must always be a non-negative whole number. As a consequence:
\begin{equation}
    \hat{H}|n\ket = \left(n + \frac{1}{2}\right)\hbar\omega|n\ket,
\end{equation}
which means that the energy eigenvalues are given by:
\begin{equation}
    E_n = \left(n + \frac{1}{2}\right)\hbar\omega.
\end{equation}

Some important relations linked to the annihilation and creation operators are:
\begin{equation}
    \hat{a} |n \ket = \sqrt{n} |n-1\ket,
\end{equation}
\begin{equation} \label{eq:qhocreaop}
    \hat{a}^\dagger |n\ket = \sqrt{n+1} |n+1\ket.
\end{equation}

These relations imply that when there is the application of the operator $\hat{a}^\dagger$ to the state $|n\ket$, and similarly of $\hat{a}$ on $|n\ket$, new states are obtained, that are also eigenstates of the number operator $\hat{N}$. These new states have their eigenvalues increased or decreased by one. This change by one corresponds to either adding or removing one unit of quantum energy, which is $\hbar\omega$. 

Applying systematically the creation operator $\hat{a}^\dagger$ to the ground state $|0\ket$ using Eq. (\ref{eq:qhocreaop}), it is possible to construct the eigenstates of both the number operator $\hat{N}$ and the Hamiltonian operator $\hat{H}$:
\begin{equation}
    |n\ket = \left[\frac{(\hat{a}^\dagger)^n}{\sqrt{n!}} \right]|0\ket.
\end{equation}
Using the orthonormality requirement for these states, the matrix elements are derived for the $\hat{a}$, $\hat{a}^\dagger$, $\hat{x}$ and $\hat{p}$ operators:
\begin{equation}
    \bra n'|\hat{a}|n\ket = \sqrt{n} \delta_{n', n-1},
\end{equation}
\begin{equation}
    \bra n'|\hat{a}^\dagger|n\ket = \sqrt{n+1} \delta_{n', n+1},
\end{equation}
\[\bra n'|\hat{x}|n \ket =\sqrt{\frac{\hbar}{2m\omega}} \times\] 
\begin{equation}
\times\left(\sqrt{n}\delta_{n',n-1} + \sqrt{n+1}\delta_{n',n+1}\right),
\end{equation}
\[\bra n'|\hat{p}|n \ket =i\sqrt{\frac{m\hbar\omega}{2}} \times\] 
\begin{equation}
\times\left(-\sqrt{n}\delta_{n',n-1} + \sqrt{n+1}\delta_{n',n+1}\right).
\end{equation}

These matrix elements allow calculating the expectation values of position and momentum operators in these states, which reveal that the ground state has a minimum uncertainty product, while excited states have larger uncertainty products. These findings satisfy the uncertainty principle and confirm the quantum behavior of the harmonic oscillator.

In addition, the following relations are also important:
\begin{equation} \label{eq:qhoxop}
    \hat{x} = \sqrt{\frac{\hbar}{2 m \omega}}(\hat{a}^\dagger + \hat{a}),
\end{equation}
\begin{equation} \label{eq:qhopop}
    \hat{p} = i\sqrt{\frac{m\hbar \omega}{2}} (\hat{a}^\dagger - \hat{a}).
\end{equation}

In the position space, the wave functions can be derived for these states. The ground state's wave function is a Gaussian, and the excited states have different wave functions with larger uncertainties, reflecting their higher energy levels. This demonstrates that the uncertainty principle holds for the harmonic oscillator and provides valuable insights into its quantum behavior. So, for the zero point energy, with $x_0 = \sqrt{\frac{\hbar}{m\omega}}$, some useful relations can be obtained:
\begin{equation}
    \big\bra \hat{x} \big\ket = \big\bra \hat{p} \big\ket = 0,
\end{equation}
\begin{equation}
    \big\bra \hat{x}^2 \big\ket = \frac{\hbar}{2m\omega} = \frac{x_0^2}{2},
\end{equation}
\begin{equation}
    \big\bra \hat{p}^2 \big\ket = \frac{\hbar m\omega}{2}.
\end{equation}

\subsubsection{Time Evolution of a Quantum Harmonic Oscillator}
For treating the time evolution of the Quantum Harmonic Oscillator, since the Hamiltonian is written in the format:
\begin{equation}
    \hat{H} = \frac{\hat{p}^2}{2m} + V(\hat{x}),
\end{equation}
it is possible to obtain the relations:
\begin{equation}\label{eq:heirel1}
    \frac{dp_i}{dt} = \frac{1}{i\hbar}[p_i, V(\hat{x})] = -\del{ }{x_i}V(\hat{x}),
\end{equation}
\begin{equation}\label{eq:heirel2}
    \frac{dx_i}{dt} = \frac{p_i}{m}.
\end{equation}
So, applying Eqs. (\ref{eq:heirel1}) and (\ref{eq:heirel2}) to the annihilation and creation operators, the following uncoupled differential equations are obtained:
\begin{equation}
    \frac{d\hat{a}^\dagger}{dt} = i\omega\hat{a}^\dagger,
\end{equation}
\begin{equation}
    \frac{d\hat{a}}{dt} = -i\omega\hat{a},
\end{equation}
with solutions:
\begin{equation}
    \hat{a}^\dagger(t) = \hat{a}^\dagger(0) e^{i\omega t},
\end{equation}
\begin{equation}
    \hat{a}(t) = \hat{a}(0) e^{-i\omega t}.
\end{equation}
These solutions imply that the operators $\hat{N}$ and $\hat{H}$ are time-independent. Moreover, using these solutions in combination with Eqs. (\ref{eq:qhoxop}) and (\ref{eq:qhopop}), the time evolution for the position and momentum is determined:
\begin{equation}
    \hat{x}(t) = \hat{x}(0)\cos\omega t + \left(\frac{\hat{p}(0)}{m\omega}\right)\sin\omega t,
\end{equation}
\begin{equation}
    \hat{p}(t) = -m\omega\hat{x}(0)\sin\omega t + \hat{p}(0)\cos\omega t.
\end{equation}

\subsubsection{Coherent States}\label{sec:coherent_states}
Defining a superposition state $\alpha$ of a Quantum Harmonic Oscillator of the two first energy levels:
\begin{equation}
    |\alpha\ket = c_0|0\ket + c_1 |1\ket.
\end{equation}
Calculating the expectation value of $x(t)$, as $\bra \alpha |\hat{x}(t)|\alpha\ket$, there is an oscillation of:
\begin{equation}
    \bra \alpha |\hat{x}(t)|\alpha\ket = 2 c_0 c_1 \sqrt{\frac{\hbar}{2 m \omega}} \cos\omega t.
\end{equation}

Given that, the expectation value of $x(t)$ exhibits oscillations. Nonetheless, an energy eigenstate does not exhibit behavior akin to a classical oscillator – in terms of oscillating expectation values for $x$ and $p$ – irrespective of how large $n$ may be. It prompts a logical inquiry: How is it possible to devise a superposition of energy eigenstates that closely emulates the classical oscillator? In the language of wave functions, it is sought a wave packet that oscillates back and forth without altering its shape. Within this framework, it is defined a coherent state, defined by the eigenvalue equation for the non-Hermitian annihilation operator $\hat{a}$:
\begin{equation}
\hat{a}|\lambda \rangle = \lambda|\lambda \rangle,
\end{equation}
where $\lambda$ generally represents a complex eigenvalue that accomplishes this task. The coherent state boasts several other noteworthy properties~\cite{sakurai_napolitano_2020}:

\begin{itemize}
    \item When expressed as a superposition of energy (or $N$) eigenstates,
    \begin{equation}
    |\lambda \rangle = \sum_{n=0}^{\infty} f(n)|n \rangle,
    \end{equation}
    the distribution of $|f(n)|^2$ relative to $n$ follows a Poisson distribution around some mean value $\bar{n}$:
    \begin{equation}
    |f(n)|^2 = \frac{{\bar{n}}^n}{n!} \exp(-\bar{n}).
    \end{equation}
    \item It can be derived by displacing the oscillator ground state by a finite distance;
    \item It maintains the minimum uncertainty product relation at all instances.
\end{itemize}

\end{document}